\documentclass{article}
\usepackage{amssymb}
\begin{document}
%
%
%
%
\newcommand{\bbbone}{{\mathchoice {\rm 1\mskip-4mu l} {\rm 1\mskip-4mu l}    {\rm 1\mskip-4.5mu l} {\rm 1\mskip-5mu l}}}
\newcommand{\bA}{{\mathbb A}}
\newcommand{\bB}{{\mathbb B}}
\newcommand{\bC}{{\mathbb C}}
\newcommand{\bD}{{\mathbb D}}
\newcommand{\bE}{{\mathbb E}}
\newcommand{\bF}{{\mathbb F}}
\newcommand{\bG}{{\mathbb G}}
\newcommand{\bH}{{\mathbb H}}
\newcommand{\bI}{{\mathbb I}}
\newcommand{\bJ}{{\mathbb J}}
\newcommand{\bK}{{\mathbb K}}
\newcommand{\bL}{{\mathbb L}}
\newcommand{\bM}{{\mathbb M}}
\newcommand{\bN}{{\mathbb N}}
\newcommand{\bO}{{\mathbb O}}
\newcommand{\bP}{{\mathbb P}}
\newcommand{\bQ}{{\mathbb Q}}
\newcommand{\bR}{{\mathbb R}}
\newcommand{\bS}{{\mathbb S}}
\newcommand{\bT}{{\mathbb T}}
\newcommand{\bU}{{\mathbb U}}
\newcommand{\bV}{{\mathbb V}}
\newcommand{\bW}{{\mathbb W}}
\newcommand{\bX}{{\mathbb X}}
\newcommand{\bY}{{\mathbb Y}}
\newcommand{\bZ}{{\mathbb Z}}
%
%
\newcommand{\al}{\alpha}
\newcommand{\be}{\beta}
\newcommand{\ga}{\gamma}
\newcommand{\de}{\delta}
\newcommand{\ep}{\epsilon}
\newcommand{\et}{\eta}
\newcommand{\ze}{\zeta}
\renewcommand{\th}{\theta} 
\newcommand{\io}{\iota}
\newcommand{\ka}{\kappa}
\newcommand{\la}{\lambda}
\newcommand{\rh}{\rho}
\newcommand{\si}{\sigma}
\newcommand{\ta}{\tau}
\newcommand{\up}{\upsilon}
\newcommand{\ph}{\phi}
\newcommand{\ch}{\chi}
\newcommand{\ps}{\psi}
\newcommand{\om}{\omega}
\newcommand{\vphi}{\varphi}
\newcommand{\vth}{\vartheta}
\newcommand{\veps}{\varepsilon}
%
%
\newcommand{\Ga}{\Gamma}
\newcommand{\De}{\Delta}
\newcommand{\Ep}{\Epsilon}
\newcommand{\Th}{\Theta}
\newcommand{\La}{\Lambda}
\newcommand{\Si}{\Sigma}
\newcommand{\Up}{\Upsilon}
\newcommand{\Ph}{\Phi}
\newcommand{\Ps}{\Psi}
\newcommand{\Om}{\Omega}
%
%
\newcommand{\bGa}{{{\rm I}\kern-.16em \Gamma}}
%
%
\newcommand{\alq}{{\bar\alpha}}
\newcommand{\beq}{{\bar\beta}}
\newcommand{\gaq}{{\bar\gamma}}
\newcommand{\deq}{{\bar\delta}}
\newcommand{\epq}{{\bar\epsilon}}
\newcommand{\etq}{{\bar\eta}}
\newcommand{\zeq}{{\bar\zeta}}
\newcommand{\thq}{{\bar\theta}}
\newcommand{\ioq}{{\bar\iota}}
\newcommand{\kaq}{{\bar\kappa}}
\newcommand{\laq}{{\bar\lambda}}
\newcommand{\rhq}{{\bar\rho}}
\newcommand{\siq}{{\bar\sigma}}
\newcommand{\taq}{{\bar\tau}}
\newcommand{\upq}{{\bar\upsilon}}
\newcommand{\phq}{{\bar\phi}}
\newcommand{\chq}{{\bar\chi}}
\newcommand{\psq}{{\bar\psi}}
\newcommand{\omq}{{\bar\omega}}
\newcommand{\vepsq}{{\bar\varepsilon}}
\newcommand{\vthq}{{\bar\vth}}
\newcommand{\vphq}{{\bar\vph}}
\newcommand{\Psq}{{\bar\Ps}}
%
%
\newcommand{\cA}{{\cal A}}
\newcommand{\cB}{{\cal B}}
\newcommand{\cC}{{\cal C}}
\newcommand{\cD}{{\cal D}}
\newcommand{\cE}{{\cal E}}
\newcommand{\cF}{{\cal F}}
\newcommand{\cG}{{\cal G}}
\newcommand{\cH}{{\cal H}}
\newcommand{\cI}{{\cal I}}
\newcommand{\cJ}{{\cal J}}
\newcommand{\cK}{{\cal K}}
\newcommand{\cL}{{\cal L}}
\newcommand{\cM}{{\cal M}}
\newcommand{\cN}{{\cal N}}
\newcommand{\cO}{{\cal O}}
\newcommand{\cP}{{\cal P}}
\newcommand{\cQ}{{\cal Q}}
\newcommand{\cR}{{\cal R}}
\newcommand{\cS}{{\cal S}}
\newcommand{\cT}{{\cal T}}
\newcommand{\cU}{{\cal U}}
\newcommand{\cV}{{\cal V}}
\newcommand{\cW}{{\cal W}}
\newcommand{\cX}{{\cal X}}
\newcommand{\cY}{{\cal Y}}
\newcommand{\cZ}{{\cal Z}}
%
%
%
\newcommand{\tG}{{\tilde G}}
\newcommand{\prP}{{\bf P}}
\newcommand{\db}{{\mkern2mu\mathchar'26\mkern-2mu\mkern-9mud}}
\newcommand{\tr}{\mbox{ tr }}
\newcommand{\del}{\partial}
\newcommand{\gtoas}[1]{{\quad\mathop{\longrightarrow}\limits_{#1}\quad}}
\newcommand{\ve}[1]{{\bf #1}}
\newcommand{\edp}[1]{e^{\prime\prime}(\ve{#1})}
\newcommand{\nato}[1]{\{ 1,\ldots,#1 \} }
\newcommand{\natz}[1]{\{ 0,\ldots,#1 \} }
\newcommand{\abs}[1]{{\left\vert #1 \right\vert}}
\newcommand{\norm}[1]{{\left\Vert #1 \right\Vert}}
\newcommand{\oo}[1]{{\mathaccent'27 #1}}
\newcommand{\Pol}[1]{\hbox{ pol}(#1)}
\newcommand{\openkrnl}[1]{\mathop{#1}\limits^{\;_\zer}}
\newcommand{\Perm}[1]{{\cal S}_{#1}}
\newcommand{\Ref}[1]{$(\ref{#1})$}
%
%
\newcommand{\True}[1]{\; \bbbone\left( #1 \right) \;}
\newcommand{\sfrac}[2]{{\textstyle \frac{#1}{#2}}}
\newcommand{\ssfrac}[2]{{\scriptstyle \frac{#1}{#2}}}
\newcommand{\vol}{{\hbox{ vol }}}
\newcommand{\dbar}{{\mkern2mu\mathchar'26\mkern-2mu\mkern-9mud}}
\newcommand{\Imp}{\Longrightarrow}
\newcommand{\Equiv}{\Longleftrightarrow}
\newcommand{\dotcup}{\mathop{\mathop{\cup}\limits^\cdot}}
\newcommand{\const}{\hbox{ \rm const }}
\newcommand{\supp}{\hbox{ \rm supp }}
\newcommand{\half}{\frac{1}{2}}
\newcommand{\ili}{\int\limits}
\newcommand{\sli}{\sum\limits}
\newcommand{\pli}{\prod\limits}
\newcommand{\lli}{\lim\limits}
\newcommand{\Laplace}{\Delta}
\newcommand{\LC}{{\Laplace_C}}
\newif\ifintrmk
%
%
\newcommand{\intremark}[1]{\ifintrmk\par\bigskip\noin\hrulefill\par\medskip \noin{\bf Internal Remark:}
#1 \hfill$\clubsuit$\par\medskip\noin\hrulefill\par\bigskip\else\fi}
\newcommand{\noin}{\noindent}
\newcommand{\nonu}{\nonumber}
\newcommand{\dst}{\displaystyle}
\newcommand{\sty}{\scriptstyle}
\newcommand{\sst}{\scriptstyle}
\newcommand{\ssst}{\scriptscriptstyle}
\newcommand{\tst}{\textstyle}



\def\suffix{ps}
\newcount\system
\global\system=3   

\def\ifundefined#1{\expandafter\ifx\csname#1\endcsname\relax}
\ifundefined{figdir}\def\figdir{}\fi
%
%
\newcount\firstline
\newdimen\pswidth  \newdimen\xleft
\newdimen\psheight \newdimen\ytop \newdimen\ybot
\newcount\justx \newcount\justy
\global\justx=0 \global\justy=0
\newdimen\vpos \newtoks\labeL 
\newread\labeLfile \newdimen\xcoord \newdimen\ycoord
\newif\ifdoit 
\newbox\labox
\newdimen\xdvikwid 
\newdimen\xdvikht
\newdimen\pspoints
\newdimen\rwi
\pspoints=1bp
\newcount\temp
\def\readdim#1{\global\read\labeLfile to \temp
\global #1=\temp pt}
%
%
%
%
\def\figcrop#1{\par
\openin\labeLfile=\figdir#1.lbl                                              
\global\read\labeLfile to\firstline\message{#1}               
\global\read\labeLfile to\temp
\readdim{\ybot}
\readdim{\xleft}
\readdim{\ytop}
\global\read\labeLfile to\justx
\global\read\labeLfile to\justy
\global\read\labeLfile to\labeL
\readdim{\pswidth}
\global\advance\pswidth by -\xleft
\readdim{\psheight}
\global\advance\ybot by -\psheight
\global\advance\psheight by -\ytop
\global\read\labeLfile to\justx
\global\read\labeLfile to\justy
\global\read\labeLfile to\labeL
\vbox to\psheight{\vfill
\ifnum\system=1
\fi 
\ifnum\system=2
\fi 
\ifnum\system=3
                                                 \fi         
\ifnum\system=4
\fi         
\ifnum\system=1
\hbox to \pswidth{\kern-\xleft\special{postscriptfile \figdir#1.\suffix }\hfil}\fi
\ifnum\system=2
\hbox to \pswidth{\kern-\xleft\special{ps: plotfile \figdir#1.\suffix }\hfil}\fi
\ifnum\system=3
\hbox to \pswidth{\kern-\xleft\includegraphics{\figdir#1.\suffix}\hfil}\fi
\ifnum\system=4
\hbox to \pswidth{\kern-\xleft\includegraphics{\figdir#1.\suffix}\hfil}\fi
\ifnum\system=5
\hbox to \pswidth{\kern-\xleft\includegraphics{\figdir#1.\suffix}\hfil}\fi 
\ifnum\system=6
   \xdvikwid=\pswidth
   \xdvikht=\psheight
   {\global\divide\xdvikwid by \pspoints}
   {\global\divide\xdvikht by \pspoints}
   \rwi=\xdvikwid
    {\global\multiply\rwi by 10}
\hbox to \pswidth{\kern-\xleft\includegraphics{\figdir#1.\suffix\space}\hfil}\fi                   
\vskip -\baselineskip
\vskip -\ybot 
\vskip-\psheight %
\hbox to\pswidth  {\hss}%
\parindent=0pt\offinterlineskip                                       
\vpos=0 pt%
\loop\readdim{\xcoord}                                 
\ifdim \xcoord < -999pt \doitfalse\else\doittrue\fi                        
\ifdoit \advance \xcoord by -\xleft
\readdim{\ycoord}
\advance \ycoord by -\ytop                              
\global\read\labeLfile to\justx                                       
\global\read\labeLfile to\justy                                       
\global\read\labeLfile to\labeL
\global\setbox\labox=\hbox{\labeL\hskip-0.3em}%
\advance\vpos by-\ycoord                                              
\vskip-\vpos \vpos=\ycoord                                         
\hbox to\pswidth{\hskip\xcoord %
\hbox to 0pt{\ifnum\justx>0\hss\fi%
\vbox to0pt{%
\ifnum\justy<2\vss\fi%
\copy\labox\kern0pt%
\ifnum\justy>0\vss\fi}%
\ifnum\justx<2\hss\fi}%
\hss}%
\repeat%
\advance\vpos by-\psheight%
\vskip-\vpos %
}\closein\labeLfile}
%
%
%
\def\figplace#1#2#3{
\openin\labeLfile=\figdir#1.lbl
\ifeof \labeLfile
       \immediate\write16{***Can't find \figdir#1.lbl; Skipping it.***}
\else  \closein\labeLfile
       \null\hskip#2\raise #3 \hbox{\figcrop{#1}}
\fi
}
%
%
%
%
\def\figput#1{
\openin\labeLfile=\figdir#1.lbl
\ifeof \labeLfile
       \immediate\write16{***Can't find \figdir#1.lbl; Skipping it.***}
\else  \closein\labeLfile
       \hbox{\figcrop{#1}}
\fi
}

\def\figdir{}
%
%
\intrmkfalse
\newcommand{\E}{{\rm e}}
\newcommand{\I}{{\rm i}}
%
%
\newtheorem{theorem}{Theorem}
\newtheorem{lemma}{Lemma}
\newenvironment{proof}{\par\noin {\it Proof:} \hspace{7pt}}%
{\hfill\hbox{\vrule width 7pt depth 0pt height 7pt}
\par\vspace{10pt}}
\newenvironment{prooff}{\par\noin {\it Proof of parts 1 and 2 of Theorem \ref{deltasatz}:} \hspace{7pt}}%
{\hfill\hbox{\vrule width 7pt depth 0pt height 7pt}
\par\vspace{10pt}}

\newcommand{\ww}{\nu}
\newcommand{\dd}{{\rm d}}
\newcommand{\pp}{{\bf p}}
\newcommand{\pcord}[1]{{\tilde #1}}
\newcommand{\rF}{{r_F}}
\newcommand{\ONE}{{\tt I}}
\newcommand{\TWO}{{\tt II}}
\newcommand{\THR}{{\tt III}}
\newcommand{\leco}{\lesssim}
\newcommand{\lenuco}{\lessapprox}
\newcommand{\estart}{{E_0}}
\newcommand{\eDeriv}{{\mathbf{D}}_e}
\newcommand{\Sc}{{\mathbf c}}
\newcommand{\Sp}{{\mathbf p}}
\newcommand{\Sq}{{\mathbf q}}
\newcommand{\Sk}{{\mathbf k}}
\newcommand{\Sn}{{\mathbf n}}
\newcommand{\St}{{\mathbf t}}
\newcommand{\SP}{{\mathbf P}}

\newcommand{\Szero}{{\mathbf 0}}
\newcommand{\Spi}{\pi\mkern-9.7mu\pi}

\newcommand{\Sx}{{\mathbf x}}
\newcommand{\Sy}{{\mathbf y}}
\newcommand{\smprod}{{\mathop{\textstyle\prod}}}
\newcommand{\smsum}{{\mathop{\textstyle\sum}}}
\newcommand{\lAng}{\left<}
\newcommand{\rAng}{\right>}
\newcommand{\squiggle}{\raisebox{2pt}{${\scriptstyle\sim}$}}
\newcommand{\randbem}[1]{\marginpar{ }}

\newcommand{\Esset}{\cE_s}
\newcommand{\Vset}{\cV}
\newcommand{\radiEset}{B_{\rm rad}}
\newcommand{\AOne}{(A1)}
\newcommand{\ATwo}{(A2)}
\newcommand{\AThr}{(A3)}
\newcommand{\AFou}{(A4)}
\newcommand{\AFiv}{(A5)}
\newcommand{\SYmm}{(Sy)}
\newcommand{\hxp}{h}
\newcommand{\FS}[1]{{S_{#1}}}
\newcommand{\gz}{{g_0}}
\newcommand{\Gz}{{G_0}}
\newcommand{\dez}{{\de_0}}
\newcommand{\waz}{{\om_0}}
\newcommand{\TV}{{\mathbf t}}
\newcommand{\NV}{{\mathbf n}}
\newcommand{\KR}{K^{(R)}}
\newcommand{\DR}{{D}}
\newcommand{\Qz}{{Q_0}}
\newcommand{\Qo}{{Q_1}}
\newcommand{\Qt}{{Q_2}}
\newcommand{\Qtt}{{Q_3}}
\newcommand{\Qf}{{Q_4}}
\newcommand{\Qff}{{Q_5}}
\newcommand{\eR}{{e^{(R)}}}
\newcommand{\laR}{{\la_R}}
\newcommand{\Cts}{C^2_s}
\newcommand{\Impeps}{\gamma}
\newcommand{\QQzer}{\tilde Q_0}
\newcommand{\QQone}{\tilde Q_1}
\newcommand{\Cvol}{Q_{\mathrm{vol}}}
\newcommand{\Umg}{\cU}
\newcommand{\uz}{u_0}
\newcommand{\Val}{\mbox{ \rm Val}\,}

\newif\ifcomments
\commentstrue
\newcommand{\Comment}[1]{\ifcomments\textsc{#1}\fi}

\begin{titlepage}

\title{An inversion theorem in Fermi surface theory \\ \null }

\author{Joel Feldman%
\thanks{ Research supported in part by the Natural Sciences and
            Engineering Research Council of Canada and by the Forschungsinstitut
            f\"ur Mathematik, ETH Z\"urich.}\\
{\small
Mathematics, UBC, Vancouver, BC, Canada\ \   V6T 1Z2}\\
{\small feldman@math.ubc.ca,\ http:/\hskip-3pt/www.math.ubc.ca/\squiggle
feldman/} \\ \\ \and  Manfred Salmhofer,  Eugene Trubowitz\\ 
{\small Mathematik,
ETH-Zentrum, 8092 Z\" urich, Switzerland }\\ {\small manfred@math.ethz.ch,
trub@math.ethz.ch} \\ {\small http:/\hskip-3pt/www.math.ethz.ch/\squiggle
manfred/ } \\ \\ }

\bigskip\bigskip

\date{December 1999}
\bigskip

\maketitle

\begin{abstract}
\noin We prove a perturbative inversion theorem for the map
between
the interacting and the noninteracting 
Fermi surface for a class of many fermion systems with  strictly
convex Fermi surfaces and  short-range interactions between the
fermions. This theorem gives a physical meaning to the counterterm
function $K$ that we use in the renormalization of these models:
$K$ can be identified as that part of the self--energy that causes
the deformation of the Fermi surface when the interaction is
turned on. 
\end{abstract}

\thispagestyle{empty}

\end{titlepage}
\section{Introduction}\label{sect1}

\noin The  Fermi surface is an important feature of the quantum
field theory of solid state models. Besides being central to the
theoretical analysis of such models it is also important from a
conceptual point of view. In experiments, one observes and
measures the Fermi surface of an 
interacting system (for brevity, we call this
the {\em interacting Fermi surface}) -- or more
precisely, an approximation to it due to positive temperature
effects, because the electrons interact with each other (say via
a screened Coulomb interaction, phonons and so on). On the other
hand, the theoretical analysis usually starts from a model of
noninteracting electrons, moving in a crystal background, which
exhibits the {\em noninteracting} Fermi surface. The effects of
the electron--electron interaction are taken into account by
`turning on a coupling constant'. Thus, while the model of
independent electrons exists only theoretically, important notions
of solid state physics, for instance Fermi liquid theory, start
from it and then incorporate the changes in the system caused by
the interaction. One of these is a change in the dispersion
relation, that gives the energy of a particle as a function of
momentum. This results in the transformation of the Fermi surface
from the noninteracting to the interacting one.

In this paper, we complete our perturbative analysis of the
regularity properties of interacting nonspherical Fermi surfaces
by proving an inversion theorem for the map between the
interacting and the free dispersion relation that we used in the
renormalization of these models. The main ingredients in the
inversion theorem are an abstract iteration theorem that
generalizes the usual contraction mapping theorem (which is not
sufficient here) and a number of regularity estimates. The
estimates are used to verify the hypotheses of this iteration
theorem. The regularity estimates are an application of the
methods and the results of \cite{FST1}, \cite{FST2}, and
\cite{FST3}, referred to as \ONE, \TWO, and \THR\ in the
following.

By `perturbative analysis' we mean that the perturbation series is
truncated at any finite order $R$ (which may be arbitrarily large)
in the coupling constant $\la$. There are situations where this
expansion can be proven to converge, so that the limit $R \to
\infty$ exists, but we do not give such bounds here.

In the remainder of this introduction, we define our class of
models and state the inversion theorem.
For a more detailed motivation, see the introductory sections of 
\ONE\ and \TWO.

\subsection{The models}
Let $\Ga$ be a nondegenerate lattice in $\bR^d$ and
\begin{equation}
\Ga^\#=\{ {\mathbf b}\in\bR^d\ :\ {\mathbf b}\cdot\ga\in
2\pi\bZ\mbox{ for all }\ga\in\Ga\}
\end{equation}
its dual lattice. We denote the first Brillouin zone by $\cB$ and
choose it to be the $d$-dimensional torus $\cB=\bR^d/\Ga^\#$. It
is compact. For example, if $\Ga = \bZ^d$, then $\Ga^\#= 2\pi
\bZ^d$ and $\cB = \bR^d/2\pi \bZ^d$. We are interested in a class
of models characterized by an action $\cA(\psi,\bar\psi)$ that is
a function of two variables $\psi=\big(\psi_{k,\si}\big)_{k\in
\bR\times\cB, \si\in\{\uparrow,\downarrow\}}$ and
$\bar\psi=\big(\bar\psi_{k,\si}\big)_{k\in \bR\times\cB,
\si\in\{\uparrow,\downarrow\}}$. Note that $\bar\psi$ is not the
complex conjugate of $\psi$. It is just another vector that is
totally independent of $\psi$. The zero component $k_0$ of $k$ is
usually thought as an energy, the final $d$ components $\Sk$ as
(crystal) momenta and $\si$ as a spin. There really should also be
a sum over a band index $n$, but it will not play a role here and
has been suppressed. In these models, the quantities one measures
are represented by other functions $f(\psi,\bar\psi)$ of the same
two vectors and the value of the observable $f(\psi,\bar\psi)$ in
the model with action $\cA(\psi,\bar\psi)$ is given formally by
the ratio of integrals
\begin{equation}
\lAng f(\psi,\bar\psi)\rAng_{\cA}={ \int\, f(\psi,\bar\psi)\,
e^{{\cal A}(\psi,\bar\psi)} \, \smprod_{k,\si}
d\psi_{k,\si}\,d\bar\psi_{k,\si} \over \int e^{{\cal
A}(\psi,\bar\psi)} \, \smprod_{k,\si}
d\psi_{k,\si}\,d\bar\psi_{k,\si}}
\end{equation}
The integrals are fermionic functional integrals. That is, linear
functionals on a Grassmann algebra.

A typical action of interest is that corresponding to a gas of
electrons, of strictly positive density, interacting through a
two--body potential $u(\Sx-\Sy)$. It is
\begin{eqnarray}
\cA_{\mu,\la}\hskip-5pt &=&-\smsum_{\si\in
\{\uparrow,\downarrow\}}
\int_{\bR\times\cB}\sfrac{d^{d+1}k}{(2\pi)^{d+1}}\
\big(ik_0-\big(\sfrac{\Sk^2}{2m}-\mu\big)\big)\bar\psi_{k,\si}\psi_{k,\si}
\nonu\\ &&- {\textstyle{\la\over 2}}
\smsum_{\si,\tau\in\{\uparrow,\downarrow\}}
\int_{\bR\times\cB}\smprod_{i=1}^4\sfrac{d^{d+1}k_i}{(2\pi)^{d+1}}
(2\pi)^{d+1} \de(k_1+k_2-k_3-k_4) \\&&\hspace{2cm}
\bar\psi_{k_1,\si}\psi_{k_3,\si}   \hat u(\Sk_1-\Sk_3)
\bar\psi_{k_2,\tau}\psi_{k_4,\tau}\nonu
\end{eqnarray}
Here $\sfrac{\Sk^2}{2m}$ is the kinetic energy of an electron,
$\mu$ is the chemical potential, which controls the density of the
gas, and $\hat u$ is the Fourier transform of the two--body
interaction. The coupling constant $\lambda$ is assumed 
to be small, so that the interaction is weak. 

More generally, when the electron gas is subject to a periodic
potential due to the crystal lattice, $\Ga$, and when the
electrons are interacting with
 the motion of the crystal lattice through the mediation of harmonic phonons,
the action is of the form
\begin{eqnarray}\label{action}
\cA_{\la}\hskip-5pt &=&-\smsum_{\si\in \{\uparrow,\downarrow\}}
\int_{\bR\times\cB}\sfrac{d^{d+1}k}{(2\pi)^{d+1}}\
\big(ik_0-E(\Sk)\big)\bar\psi_{k,\si}\psi_{k,\si} \nonu\\ &&-
{\textstyle{\la\over 2}}
\smsum_{\si,\tau\in\{\uparrow,\downarrow\}}
\int_{\bR\times\cB}\smprod_{i=1}^4\sfrac{d^{d+1}k_i}{(2\pi)^{d+1}}
(2\pi)^{d+1} \de(k_1+k_2-k_3-k_4) \\&&\hspace{2 cm}
\bar\psi_{k_1,\si}\psi_{k_3,\si}   \hat
v(k_{1,0}-k_{3,0},\Sk_1-\Sk_3)
\bar\psi_{k_2,\tau}\psi_{k_4,\tau}\nonu
\end{eqnarray}
where $E(\Sk)$ is the dispersion relation minus the chemical
potential $\mu$.
\subsection{The class of dispersion relations}\label{dispclassect}
Let $\cF$ be a  fundamental cell for the action of the translation
group $\Ga^\#$. In other words, $\cF$ is an open set in $\bR^d$
with the property that it together with its translates under
$\Ga^\#$ are dense in $\bR^d$. For example, if $\Ga = \bZ^d$, then
$\Ga^\#= 2\pi \bZ^d$
 and we may choose $\cF = (-\pi,\pi)^d$.
Let $\cF_2=\{\Sp\in \cF: 2\Sp \in \cF\}$. For a continuous
function $E$ from $\cB$ to $\bR$ let
\begin{equation}
\FS{E} = \{ \Sp \in \cB: E(\Sp)=0\}
\end{equation}
be the corresponding Fermi surface and $\cI_E=\{ \Sp \in \cB:
E(\Sp)<0\}$ the corresponding Fermi sea. For $k \ge 2$ let
\begin{equation}
C_s^k(\cB,\bR) = \{ E \in C^k(\cB,\bR):  E(-\Sp)=E(\Sp)\mbox{ for
all } \Sp \in \cB\}
\end{equation}
 With the norm
$|f|_k = \sum_{|\al | \le k} \norm{D^\al f}_\infty$, it is a
Banach space. For $E\in C_s^k(\cB,\bR)$, let $B_\veps^{(k)} (E) =
\{ e\in C^k_s(\cB,\bR): \abs{e-E}_k < \veps\}$.

For positive constants $\dez,\gz,\Gz,\waz$, let $\Esset=\Esset
(\dez,\gz,\Gz,\waz )$ be the set of all $E\in\Cts (\cB,\bR)$ that
satisfy the following conditions
\begin{eqnarray}
(i)& &\FS{E} \subset \cF_2,\ \cI_E \ne \emptyset,\ \cI_E \ne \cB,
\ d(\FS{E},\del \cF_2) > \dez,  \nonu\\ (ii)& &\abs{\nabla E(\Sp)}
> \gz\mbox{ for all } \Sp \in \FS{E}:  \nonu\\ (iii)& & \abs{E}_2
< \Gz \nonu\\ (iv)& & (\TV (\Sp) , E''(\Sp) \TV(\Sp)) > \waz
\mbox{ for all }\Sp\in \FS{E} \mbox{ and all unit vectors }
\TV(\Sp)\nonu\\ & &\hspace{1in} \mbox{ tangent to }\FS{E}\mbox{ at
}\Sp\nonu
\end{eqnarray}
Since $E$ is $C^2$, the condition that $\nabla E\ne 0$ on $\FS{E}$
implies that the Fermi surface $\FS{E}$ is a $(d-1)$--dimensional
$C^2$-submanifold of $\cB$, (in $d=2$, the `surface' is a curve).
The condition $(\TV (\Sp) , E''(\Sp) \TV(\Sp)) > \waz$ implies
that $\FS{E}$ has strictly positive curvature everywhere.

\bigskip\noin
The set $\Esset$ is open in $(C_s^2(\cB,\bR), \abs{\;\cdot\;}_2)$.
In this paper, we fix any $\dez>0,\ \gz>0,\ \waz>0$ and
$\Gz>\max\{\gz,\waz\}$.

\subsection{The class of interactions}
We also define the class $\Vset$ of allowed interactions to be the
set of all functions $V$, whose Fourier transforms $\hat
v(p_0,\Sp)$ obey
\begin{eqnarray}
(i)& & \big|\hat v\big|_2 \le 1 \nonu\\ 
(ii)& & \hat v(-p_0 , \Sp)
= \overline{\hat v(p_0,\Sp)}\nonu\\ 
(iii)& & \hat v(p_0 , -\Sp) =
\hat v(p_0,\Sp)\nonu\\ 
(iv)& & \mbox{There is a bounded
function $\tilde v \in C^2 (\cB, \bR)$ and an $\al > 0$ such that}
\nonu\\ & &\hspace{1.5in} \limsup_{p_0 \to \infty}\
|p_0|^\al\sup_{\Sp}\big|\hat v(p_0,\Sp) - \tilde
v(\Sp)\big|<\infty\nonu
\end{eqnarray}
Condition (iv) is used only in the large $k_0$ regime. If an
ultraviolet cutoff is placed on $k_0$, it may be omitted.
Condition (i) implies that the interaction in momentum space,
$\hat v$, is in $C^2(\bR^{d+1})$. This is the case if the position
space integral kernel $V(x-y)$ is bounded by
$\sfrac{\const}{1+\abs{x-y}^{d+3+\veps}}$ for some $\veps>0$. The
$1$ in the condition $\abs{\hat v}_2 \le 1$ is not a restriction, 
since $V$ and $\la $ appear only in the combination $\la V$ in the
definition of the model, so a rescaling of $V$ can be absorbed by
a rescaling of $\la$.

\subsection{The counterterm function}
In \ONE, we constructed a counterterm function $K$ as a formal
power series in $\la$,
\begin{equation}
K(e,\la V, \Sp) = \sli_{r =1}^\infty \la^r K_r (e,V, \Sp)
\end{equation}
where $K_r: \cD \times\cV\times \cB \to \bR$ is defined for 
a set $\cD$ of dispersion relations $e$ with $\Esset \subset \cD$.
The conditions required for having a finite $K_r$ for all $r$
are much weaker than the conditions we impose here
(see \ONE\ and Section \ref{sect4}).
$K$ is constructed such that, for a model with action
\begin{eqnarray}\label{actionK}
\cA_{\la}\hskip-5pt &=&-\smsum_{\si} \int_{\bR\times\cB}
\sfrac{d^{d+1}k}{(2\pi)^{d+1}}\  \big(ik_0-e(\Sk)-K(e,\la V,
\Sk)\big)\bar\psi_{k,\si}\psi_{k,\si} \nonu\\ &&-
{\textstyle{\la\over 2}} \smsum_{\si,\tau}
\int_{\bR\times\cB}\smprod_{i=1}^4\sfrac{d^{d+1}k_i}{(2\pi)^{d+1}}
(2\pi)^{d+1} \de(k_1+k_2-k_3-k_4) \\&&\hspace{2 cm}
\bar\psi_{k_1,\si}\psi_{k_3,\si}   \hat
v(k_{1,0}-k_{3,0},\Sk_1-\Sk_3)
\bar\psi_{k_2,\tau}\psi_{k_4,\tau}
\nonu
\end{eqnarray}
 the Fermi surface of the interacting model
is fixed to $\FS{e}$, independently of $\la$. The function
$K(\Sp)$ is real--valued, and under the symmetry hypotheses made
here, $K(-\Sp) = K(\Sp)$. By introducing the counterterm function,
we removed the infrared divergences to all orders in the
perturbation expansion in powers of $\la$. That is, when the
expansion is truncated at any finite order $R$, all Green
functions are finite almost everywhere. We showed in \ONE\ that
the counterterm function to any order $R$ in $\la$,
\begin{equation}
\KR (e, \la V) = \sli_{r=1}^R \la^r K_r(e, V),
\end{equation}
is differentiable in $\Sp$ and $e$ (and, of course, $C^\infty$ in
$\la$ since it is a polynomial for any finite $R$).

Thus a model that has an action  whose quartic part (in the
fields) is that corresponding to $V$ and whose quadratic part is
that corresponding to a dispersion relation $E$ will have an
interacting Fermi surface that is the zero set of a dispersion
relation $e$ if
\begin{equation}
e+K (e,\la V) = E.
\end{equation}
In this paper we take a given $E$ and $V$ and solve
\begin{equation}\label{WltfRml}
e+\KR (e,\la V) = E.
\end{equation}
for $e=\eR(E,\la V)$.
The dispersion relation $e$ that appears in the propagator is only
an auxiliary quantity, which is to be determined by \Ref{WltfRml}.
We shall solve \Ref{WltfRml} by iteration, starting from the given
$E$. Clearly this requires having bounds with uniform constants on
a set of dispersion relations that is mapped to itself by the
function $\bbbone+\KR$.

We proved in \ONE--\THR\ that the following estimate holds
(Theorem \THR.3.13). 
For all $r \ge 1$, there are constants $\ka_r > 0$ such that, 
for all $e \in\Esset(\dez,\gz,\Gz,\waz)$ and $V\in\Vset$, 
the contribution $K_r (e, V)$ is in $\Cts (\cB,\bR)$ and obeys
\begin{equation}\label{karest}
\abs{K_r (e, V)}_2 \le \ka_r .\end{equation} The constant $\ka_r$
depends only on $(\dez,\gz,\Gz,\waz )$ and $r$. Consequently,
$\KR$ satisfies
\begin{equation}\label{q1}
\abs{\KR (e, \la V)}_2 \le \sli_{r=1}^R |\la|^r \ka_r
\end{equation}
so $|\KR (e, \la V)|_2$ can be made arbitrarily small by
decreasing $\la$. Because $\Esset$ is open in $\abs{\;\cdot\;}_2$,
$e+\KR(e,\la V) \in \Esset$ if $e \in \Esset$ and $\la$ is sufficiently
small. 
\subsection{The inversion theorem}
To show that an iteration scheme for the solution converges, we
need to have bounds for the distance between successive elements
of the iteration sequence. For technical reasons that have nothing
to do with the analysis of \ONE--\THR\ and that will be explained
later, we have to restrict to  dispersion relations that have
certain third order derivatives bounded, in order to control the
distance between successive iterates. This is the reason why, in
the following theorem, the starting $\estart$ is required to be in
$C^3$.
\begin{theorem} \label{invsatz}
Let $\dez,\gz,\waz>0$, $\Gz>\max\{\gz,\waz\}$ and $R \in \bN$.
Then there is a $\laR > 0$ such that for each $\abs{\la}\le \laR$,
each $E \in \Esset (\dez,\gz,\Gz,\waz) \cap C^3(\cB,\bR)$ and each
$V \in \Vset$, there is a unique $\eR\in\Esset
(\dez/2,\gz/2,2\Gz,\waz/2 )$ solving \Ref{WltfRml}. Moreover,
there is a constant $A_R > 0$ such that
\begin{equation}\label{invsimpl}
\abs{\eR -E}_2 \le A_R \, \abs{\la}.
\end{equation}
\end{theorem}
Theorem \ref{invsatz} follows from the more detailed Theorem
\ref{deltasatz} below. 
We shall discuss the more detailed theorems about inversion
in Section \ref{dissect}. 

In this paper, we do not prove optimal bounds about the
$R$--dependence of $\laR$. For the models at hand, in particular
because of the symmetry $E(\Sp)=E(-\Sp)$, one expects that
convergence does not hold at zero temperature. That is, one
expects $\laR \to 0$ as $R \to \infty$.
The reason for this is that at temperatures below a critical
temperature, the ground state of the system is superconducting, 
in which case the above perturbation expansion cannot converge.  
As noted in \cite{crg}, at
a positive temperature $T = \frac{1}{\be} > 0$, one can expect
convergence of the expansion for coupling constants $\la $ in the
region where $\la \log \be$ is small enough, that is, 
for $T \ge T_0 \E^{-\la_0/|\la|}$ where $\la_0$ and 
$T_0$ are fixed constants (see \cite{crg} for a
Fermi liquid criterion based on this convergence). For $d=2$, a
proof of this may be possible using the techniques of \cite{FMRT}.
The bounds derived here do not change in an essential way at
positive temperature. So a variant of our theorems can be expected
to hold in this convergent positive temperature regime. Note,
however that convergence of the expansion for $K$ does not imply
that the solution of the inversion equation can be expanded in
$\la$. In fact, it can't. See \cite{S,Brisbane} for an informal
explanation.

\section{Preliminaries}\label{sect2}

\subsection{Coordinates}
Since $e$ is going to change under the iteration, it is convenient
to use momentum space coordinates that are independent of $e$.
Under our assumptions, we can simply use polar coordinates
in addition to the Fermi surface coordinates that we used in
\ONE--\THR. We shall review the latter shortly. It will be
important that the angular variables $\th$ are the same in both
coordinate systems. Only the radial coordinate is different.

\medskip\noin
{\bf Polar coordinates:} Consider a small ball $B$ around an
$\estart \in \Esset$. Regard a small neighbourhood of the Fermi
surface $S_\estart$ 
as a subset of $\bR^d$ instead of the torus $\cB$
and introduce polar coordinates $(r,\th) \in \bR_0^+ \times
S^{d-1}$, $\Sp = \pp(r,\th)$.     For $d=2$, $\th \in S^1$. In
polar coordinates, the Fermi surface can be parametrized, for $e
\in B$, as
\begin{equation}
S_e =\{ \pp(\rF (e,\th), \th): \th \in S^{d-1}\}
\end{equation}
with $\rF: B \times S^{d-1} \to \bR^+$. If $e \in C^k (\cB,\bR)$,
then $\rF \in C^k (S^{d-1},\bR^+)$. 

\begin{lemma}\label{convsurface}
Let $0<k\le K$. Let $S$ be a $(d-1)$--dimensional $C^2$ convex surface 
in $\bR^d$ all 
of whose principal curvatures are between $k$ and $K$. Let $\Sc_1,\Sc_2$
be any two maximally separated points of $S$. That is, $\Sc_1,\Sc_2\in S$
with
\begin{equation}
\|\Sc_1-\Sc_2\|=\max\big\{\|\Sp_1-\Sp_2\|\ :\ \Sp_1,\Sp_2\in S\big\}
\end{equation}
Set $\Sc=\half\big(\Sc_1+\Sc_2\big)$.
 Then, for every $\Sp\in S$,
\begin{equation}\label{convradius}
\sfrac{1}{K}\le\|\Sp-\Sc\|\le\sfrac{1}{k}
\end{equation}
and the angle $\th(\Sp)$ between $\Sp-\Sc$ and the outward pointing
 normal vector $\Sn(\Sp)$ to $S$ at $\Sp$ obeys
\begin{equation}\label{convangle}
\cos(\th(\Sp))\ge\sfrac{k}{K}
\end{equation}
If, in addition, $-\Sp\in S$ for every $\Sp\in S$, then $\Sc$ is the origin.
\end{lemma}

\begin{proof} 
See Appendix \ref{convexproof}
\end{proof}

\begin{lemma}\label{radderiv}
Let $\dez,\gz,\waz>0$ and $\Gz>\max\{\gz,\waz\}$. There are
$\veps,\ r_0,\ g_1>0$ such that, for every $\estart\in\Esset
(\dez,\gz,\Gz,\waz )$ and every $e\in B_\veps^{(2)} (\estart)$
\begin{equation}
e\in\Esset (\dez/2,\gz/2,2\Gz,\waz/2 )
\end{equation}
and
\begin{eqnarray}
\!\!\!\!\!
&\big|\rF(e,\th)-\rF(\estart,\th)\big|\le r_0 &\mbox{ for all }\ \
\th\in S^{d-1}\label{pfs}\\
&\frac{\del}{\del r} e(\pp(r,\th)) > g_1 &\mbox{ for all }\ \
\big|r-\rF(\estart,\th)\big|\le 2r_0,\ \th\in S^{d-1} \label{rg0}
\end{eqnarray}
Note that the constants $\veps,\ r_0$ and $g_1$ are independent of
$\estart$.
\end{lemma}
\begin{proof} 
See Appendix \ref{radproof}
\end{proof}

\noindent
Let $\estart,\ r_0$ and $\veps$ be as in Lemma \ref{radderiv}. 
Set
\begin{eqnarray}
\underline{R}(\th)&=&\rF(\estart,\th)-2r_0\nonu\\
\overline{R}(\th)&=&\rF(\estart,\th)+2r_0
\end{eqnarray}
and
\begin{eqnarray}\label{Ringerl}
A&=&\{ (r,\th): \th \in S^{d-1}, \underline{R}(\th) < r <
\overline{R}(\th)\}\nonu\\ \tilde A &=& \{\pp(r,\th): (r,\th) \in
A\}
\end{eqnarray}
Then, $S_e = \{ \Sp : e(\Sp) = 0\} \subset \tilde A$ for all $e\in
B_\veps^{(2)}(\estart)$.

We use the notation $\pcord{F} (r,\th) = F(\pp(r,\th))$ for
functions in terms of the variables $r$ and $\th$, for instance,
$\pcord{e} (r,\th) = e(\pp(r,\th))$. The above Lemma then states
that for all $(r,\th) \in A$, and all $e\in B_\veps^{(2)}
(\estart)$,
 $\del_r \pcord{e} (r,\th)>  g_1 > 0$.

We could have introduced coordinates in the annulus $\tilde A$,
based on any vector field that is transversal to $S_{\estart}$.
This would only have changed the constant in the lower bound for
$\del _r e$.

\medskip\noin
{\bf Fermi surface coordinates:} These are the coordinates used in
\ONE--\THR. They are the polar coordinate $\th$ and $\rh= e(\Sp)$,
and thus obviously depend on $e$. We denote the corresponding
inverse map, whose range is a neighbourhood of $\estart$'s Fermi
surface, by  $\Spi$:
\begin{equation}
\Spi: (-\rh_0,\rh_0) \times S^{d-1} \to \cB, \quad (\rh,\th) \mapsto
\Spi(\rh,\th)
\end{equation}
Clearly $e(\Spi(\rh,\th))=\rh$.

The projection to the Fermi surface is obtained by setting
$\rh=0$. In terms of the polar coordinates, it is constructed as
follows. If $\pcord{F}: B_\veps^{(2)}(\estart) \times A \to \bC$
maps $(e,r,\th) \mapsto \pcord{F}(e,r,\th)$, then $\ell_e
\pcord{F} (e,r,\th) = \pcord{F}(e,\rF(e,\th),\th)$. Obviously,
$\del_r (\ell_e F)=0$. Observe that $\Spi (0,\th) =
\pp(\rF(e,\th),\th)$.

\subsection{Norms}
Let $\norm{\;\cdot\;}_k$ be the seminorm
$\norm{F}_k = \sli_{|\al|=k} \sup\limits_{p}\abs{\del^\al F(p)}$
and
\begin{equation}
\abs{F}_k = \sli_{l=0}^k \norm{F}_l     .
\end{equation}
It does not matter whether we use the norm in Cartesian or polar
coordinates since the two are equivalent.

We define the {\em radial} norms for $p \ge 1$ as
\begin{equation}
\abs{F}_{p,r} = \abs{F}_{p-1} + \norm{\del_r \pcord{F}}_{p-1}
\end{equation}
and denote the angular norms for $p \ge 0$ as $\abs{F}_{p,\th}$.
In the latter norms, all derivatives are taken in the
$\th$--directions.
\randbem{where used?}
\begin{lemma}
\begin{enumerate}
\item
$\abs{F}_{p,r} \le \abs{F}_{p+1}$.

\item
For all $e \in B_\veps^{(2)}(\estart)$, $\del_r \ell_e F =0$, and
\begin{equation}
\abs{\ell_e F}_{p,r} =\abs{\ell_e F}_{p-1} = \abs{\ell_e
F}_{p-1,\th}.
\end{equation}

\item
\begin{eqnarray}
\abs{FG}_p &\le& 2^p \; \abs{F}_p \; \abs{G}_p, \label{FGp}\\
\abs{FG}_p &\le& \norm{F}_0 \norm{G}_p + \norm{F}_p \norm{G}_0 +
2^{p+1} \abs{F}_{p-1} \abs{G}_{p-1},\nonu
\end{eqnarray}

\item
\begin{eqnarray}
\abs{FG}_{p+1,r} & \le & 2^{p+2} \abs{F}_p \abs{G}_p +
\norm{\del_r F}_{p} \norm{G}_0 + \norm{F}_0 \norm{\del_r G}_p
\nonu\\ &\le& 2^{p+2} \left( \abs{F}_{p+1,r} \abs{G}_p + \abs{F}_p
\abs{G}_{p+1,r} \right).
\end{eqnarray}
\end{enumerate}
\end{lemma}

\begin{proof}
The first statement is an immediate consequence of $\norm{\del_r
F}_p \le \norm{F}_{p+1}$. The second statement is an immediate
consequence of the observation that the localization map $\ell_e$
does not depend on $r$. For the third and fourth statements, use
the Leibniz rule
 and that $\prod {\al_k \choose \be_k} \le {p \choose q}$
for all $\al_1 + \ldots + \al_n=p$ and $\be_1 + \ldots +\be_n= q$
(all nonnegative),
to prove that
\begin{equation}
\norm{FG}_p \le \sli_{q=0}^p {p \choose q} \norm{F}_q
\norm{G}_{p-q}
\end{equation}
and
\begin{equation}
\norm{\del_r (FG)}_p \le \sli_{q=0}^p {p \choose q} \left(
\norm{\del_r F}_q \norm{G}_{p-q} + \norm{F}_{p-q} \norm{\del_r
G}_q \right) .
\end{equation}
\end{proof}

\newpage

\section{The iteration}\label{sect3}

Given $e_0$, $e_1$, and $t \in [0,1]$, denote $e_t = (1-t) e_0 + t
e_1$.

\begin{theorem}\label{deltasatz}
Let $\dez,\gz,\waz>0$ and $\Gz>\max\{\gz,\waz\}$. Let $\veps>0$ be
as in Lemma \ref{radderiv}.
\begin{enumerate}
\item
{\bf Regularity. }
For each $R\in \bN$, there is a constant $\DR \ge 1$  such that
for all $\abs{\la} \le 1$,
 all $V \in \Vset$ and all $e\in\Esset (\dez/2,\gz/2,2\Gz,\waz/2 )$
\begin{equation}\label{r1}
\abs{\KR (e) }_{3,r} = \abs{\KR (e) }_2 < \DR \abs{\la} .
\end{equation}

\item
{\bf Norm bounds for the iteration. }
There is $0< \de < 1$ (independent of $\dez,\gz,\Gz,\waz$) and,
for each $ R \in \bN$, there are constants $\Qz,\Qo \ge 1$ such
that for all $\abs{\la} \le 1$, all $V \in \Vset$, all $E\in\Esset
(\dez,\gz,\Gz,\waz)$
 and all $e_0$ and $e_1\in B_\veps^{(2)} (E) \cap C^3$
\begin{eqnarray}\label{dr2}
\abs{\KR (e_1) - \KR(e_0)}_0 &\le& \Qz \abs{\la}\; \abs{e_1-e_0}_0
\label{dr0}\\
\abs{\KR (e_1) - \KR(e_0)}_1 &\le& \Qz \abs{\la}\Big[
\abs{e_1-e_0}_0^\de+ \abs{e_1-e_0}_1\Big] \label{dr1}\\
\abs{\KR (e_1) - \KR(e_0)}_{3,r} &\le& \Qz \abs{\la}\Big[
\abs{e_1-e_0}_1^\de+ \abs{e_1-e_0}_2\Big]\nonu\\ &+&
\Qo \abs{\la}\;\sup\limits_{0\le t\le 1}
\abs{e_t}_{3,r}\; \abs{e_1-e_0}_0 . \label{eq35}
\end{eqnarray}
In addition, if $V_1,V_2 \in \Vset$ with $\abs{V_1-V_2}_2\le 1$
and $ e \in \Esset$ 
then
\begin{equation}\label{dr3}
\abs{\KR (e,\la V_1) - \KR(e,\la
V_2)}_2\le\Qz\abs{\la}\;\abs{V_1-V_2}_2 .
\end{equation}

\item
{\bf Existence of a unique solution to the inversion equation. }
Let $E\in\Esset (\dez,\gz,\Gz,\waz)$ with $\abs{E}_{3,r} = G_3 <
\infty$. (This is the case if, for example, $E \in \Esset \cap
C^3$). Set $Q=\max\{\Qz+\Qo(1+G_3),D\}$. 
Let  
\begin{equation}\label{radidef}
\radiEset = \{ e\in\Esset (\dez/2,\gz/2,2\Gz,\waz/2 ):
\abs{e-E}_2 < \veps, \abs{e-E}_{3,r} < 1\}
\end{equation}
and let $\laR> 0$ be such that $Q\laR <\min\{1,\veps\}$. 
Then for all $\abs{\la} \le\laR$ and all $V \in \Vset$, 
there is a unique $\eR\in\radiEset$ such that 
$E=\eR+\KR (\eR ,\la V)$. 
Moreover
\begin{equation}\label{Ala}
\abs{\eR -E}_{3,r} \le \DR\; \abs{\la} .
\end{equation}

\item
{\bf Continuity in $E$ and $V$. }
Let $E,E'\in\Esset(\dez,\gz,\Gz,\waz)$ satisfy 
$\abs{E}_{3,r}$, $\abs{E'}_{3,r}\le G_3$ and
$\abs{E-E'}_{3,r}<\veps/2$. Then, for all $\abs{\la}\le\laR/2$
and all $V,V' \in \Vset$ with $\abs{V-V'}_2\le 1$,
\begin{eqnarray}\label{eq39}
&&
\abs{\eR (E ,\la V) - \eR (E' ,\la V')}_2 
\nonu\\
&&\quad\le
4\Big( \abs{E-E'}_2+\abs{E-E'}_1^\de+\abs{E-E'}_0^{\de^2}
+\abs{V-V'}_2^{\de^2}\Big) .
\end{eqnarray}

\end{enumerate}
\end{theorem}

\begin{proof}
Part 1 was proven in \ONE--\THR: equation \Ref{r1} follows
directly from \Ref{q1}. The bound \Ref{dr0} follows from Theorem
\ONE.3.5 by summation over $r \in \{ 1, \ldots R\}$. We reexplain
that argument briefly in the proof of Theorem \ref{oisonacha}
(Section \ref{th4proof}). 
We shall shortly prove the remaining statements of part 2 
from the more
detailed estimates given in Theorem \ref{skalensatz}.

\vspace{.25cm} 
\noindent
To prove part 3, fix $R$, let $V \in \Vset$, and
denote for brevity $K(E)=\KR (E,\la V)$ and $B=\radiEset$. Define
$\Ph : B \to \Cts (\cB,\bR)$ by
\begin{equation}\label{Phdef}
\Ph (e) = E - K(e).
\end{equation}
By \Ref{r1} and the hypothesis on $\laR$,
\begin{equation}\label{welldef}
|\Ph(e)-E|_{3,r} = |K(e)|_{3,r}\le \DR \abs{\la} \le Q \laR <
\min\{\veps,1\} < \veps
\end{equation}
so $\Ph(B) \subset B$. Thus the sequence $(e_n)_{n \ge 0}$ given
by $e_0=E$ and $e_{n+1}=\Ph(e_n)$ is well-defined. For $n \ge 1$,
let $f_n = e_n-e_{n-1}$. Then $f_1= - K(E)$, $e_n=E+\sum_{k=1}^n
f_k$, and
\begin{equation}\label{kphi}
f_{n+1} = \Ph(e_n)-\Ph(e_{n-1}) = K(e_{n-1}) - K(e_n)
.\end{equation} 
Let $|\la| \le \laR$.
We show that, for all $n \ge 1$,
\begin{eqnarray}
\abs{f_n}_0 &\le& (Q \abs{\la})^n \label{ih0}\\ \abs{f_n}_1 &\le&
B_R (\la )\;(Q \abs{\la})^{n\de}  \label{ih1}\\ \abs{f_n}_{3,r}
&\le& C_R (\la )\;\max\{B_R (\la )^\de,1\}\; (Q
\abs{\la})^{n\de^2}\label{ih3}
\end{eqnarray}
with
\begin{equation}\label{CRdef}
B_R(\la ) = \frac{(Q\abs{\la})^{1-\de}}{1-(Q
\abs{\la})^{1-\de}},\hspace{1cm} C_R(\la ) =
\frac{(Q\abs{\la})^{1-\de^2}}{1-(Q \abs{\la})^{1-\de^2}}
.\end{equation} Once this is done, \Ref{ih3} implies that $\sum
f_n$ converges in $\abs{\; \cdot\; }_{3,r}$. Thus $\eR =
\lim_{n\to \infty} e_n$ exists. By \Ref{eq35}, $\Ph$ is continuous
in $\abs{\; \cdot\; }_{3,r}$, so $\Ph(\eR)=\eR$ and hence, by
\Ref{Phdef}, $E=\eR+K(\eR)$. By \Ref{welldef}, every $e_n$ obeys
$|e_n-E|_{3,r}\le \DR \abs{\la} $, so $\eR$ satisfies \Ref{Ala}.
Since $\Qz\abs{\la}<1$, uniqueness follows from \Ref{dr0}.

We prove \Ref{ih0}--\Ref{CRdef} by induction on $n$. The statements
are true for $n=1$ because
\begin{equation}
\abs{f_1}_0 \le \abs{f_1}_1 \le \abs{f_1}_{3,r} = \abs{K(E)}_{3,r}
\le \DR \abs{\la}\le Q\abs{\la}
\end{equation}
and
\begin{eqnarray}
(Q \abs{\la})^{\de}B_R(\la ) 
&=& 
\frac{Q\abs{\la}}{1-(Q
\abs{\la})^{1-\de}}>Q\abs{\la},
\nonu\\
(Q \abs{\la})^{\de^2}C_R(\la ) 
&=& 
\frac{Q\abs{\la}}{1-(Q
\abs{\la})^{1-\de^2}}>Q\abs{\la} .
\end{eqnarray}
Assume \Ref{ih0}--\Ref{CRdef} to hold for $n$. 
By \Ref{kphi}, \Ref{dr0},
and the inductive hypothesis \Ref{ih0},
\begin{equation}
\abs{f_{n+1}}_0 = \abs{K(e_n)-K(e_{n-1})}_0 \le \Qz\abs{\la}\,
|f_n|_0 \le \Qz\abs{\la}  \, (Q \abs{\la})^n\le (Q\abs{\la})^{n+1}
\end{equation}
which proves \Ref{ih0} for $n+1$.

By \Ref{kphi}, \Ref{dr1},  \Ref{ih0} and the inductive hypothesis
\Ref{ih1},
\begin{eqnarray}
\abs{f_{n+1}}_1 &=& \abs{K(e_n)-K(e_{n-1})}_1 \nonu\\ &\le& \Qz
\abs{\la} \;\Big[|f_n|_0^\de +  |f_n|_1\Big] \nonu\\ &\le& Q
\abs{\la} \;\Big[(Q\abs{\la})^{n\de} +
B_R(\la)\,(Q\abs{\la})^{n\de}\Big] \nonu\\ &=&
\Big[(Q\abs{\la})^{1-\de} +
B_R(\la)\;(Q\abs{\la})^{1-\de}\Big]\,(Q\abs{\la})^{(n+1)\de} 
\end{eqnarray}
Thus the induction goes through for \Ref{ih1} if
\begin{equation}\label{e24}
(Q\abs{\la})^{1-\de} +  B_R(\la)\;(Q\abs{\la})^{1-\de} \le
B_R(\la)
\end{equation}
With the definition \Ref{CRdef}, equality holds in \Ref{e24}.

By \Ref{kphi}, \Ref{dr2}, \Ref{ih1} and the inductive hypothesis
\Ref{ih3},
\begin{eqnarray}
\abs{f_{n+1}}_{3,r} &=& \abs{K(e_n)-K(e_{n-1})}_{3,r}
=\abs{K(e_n)-K(e_{n-1})}_2 
\nonu\\ 
&\le&
\Qz \abs{\la} |f_n|_1^\de
+ \Qz \abs{\la} |f_n|_2 + \Qo (1+G_3) \abs{\la} |f_n|_0 
\nonu\\
&\le& 
Q \abs{\la} \;\Big[|f_n|_1^\de +  |f_n|_2\Big] 
\\ 
&\le&
Q \abs{\la} \;\Big[B_R(\la)^\de\;(Q\abs{\la})^{n\de^2} +
C_R(\la)\max\{B_R(\la)^\de,1\}\;(Q\abs{\la})^{n\de^2}\Big] 
\nonu\\
&\le& 
\max\{B_R(\la)^\de,1\}\Big[ (Q\abs{\la})^{1-\de^2}+
(Q\abs{\la})^{1-\de^2}\;C_R(\la)\Big]\;(Q\abs{\la})^{(n+1)\de^2}
\nonu
\end{eqnarray}
Here we used that $e_n \in B$ implies 
$|e_n|_{3,r} \le 1+ |E|_{3,r} \le 1+G_3$.
Thus the induction goes through for \Ref{ih3} if
\begin{equation}\label{e25}
(Q\abs{\la})^{1-\de^2}+  (Q\abs{\la})^{1-\de^2}\;C_R(\la)\le
C_R(\la)
\end{equation}
With the definition \Ref{CRdef}, equality holds in \Ref{e25}. This
completes the proof of part 3.

We now prove part 4. Denote for brevity
$e=\eR(E,\la V),\ e'=\eR(E',\la V')$ and $K=\KR$. 
First, observe
that both $e,e'\in \radiEset \subset B_\veps^{(2)}(E)$ 
because, by part 3,
\begin{eqnarray}
\abs{e-E}_2&\le& D\abs{\la}\le D\laR/2<\veps/2\nonu\\
\abs{e'-E}_2&\le&\abs{e'-E'}_2+\abs{E-E'}_2\le
D\laR/2+\veps/2<\veps,
\end{eqnarray}
and because $\abs{e-E}_{3,r} \le D |\la| < 1$ and 
$\abs{e'-E}_{3,r} \le D |\la| +\veps/2 < 1$ hold by \Ref{Ala}.
Thus $\max\{|e|_{3,r},|e'|_{3,r}\} \le 1+G_3$. 

By definition, $e$ and $e'$ obey
$E=e+K(e,\la V)$ and $E'=e'+K(e',\la V')$.
Hence
\begin{eqnarray}
E-E'
&=&
e-e'+K(e,\la V)-K(e',\la V')
\\ 
&=&
e-e'+K(e,\la V)-K(e',\la V)+K(e',\la V)-K(e',\la V')
\nonu
\end{eqnarray}
so that, by \Ref{dr0} and \Ref{dr3},
\begin{equation}
\abs{e-e'}_0\le\abs{E-E'}_0+\Qz \abs{\la}\; \abs{e-e'}_0 +\Qz
\abs{\la}\; \abs{V-V'}_2
\end{equation}
Recalling that $\Qz\abs{\la}\le \half Q\laR<\half$,
\begin{equation}
\abs{e-e'}_0 \le 2\Big( \abs{E-E'}_0+\sfrac{1}{2}\;
\abs{V-V'}_2\Big) \le 2\abs{E-E'}_0+ \abs{V-V'}_2
\end{equation}
Similarly, by \Ref{dr1} and \Ref{dr3},
\begin{equation}
\abs{e-e'}_1\le\abs{E-E'}_1+\Qz \abs{\la}\;
 \Big[\abs{e-e'}_0^\de+\abs{e-e'}_1\Big]
+\Qz \abs{\la}\; \abs{V-V'}_2
\end{equation}
and
\begin{eqnarray}
\abs{e-e'}_1&\le&2\Big( \abs{E-E'}_1
+\sfrac{1}{2}\;\abs{e-e'}_0^\de+\sfrac{1}{2}\;
\abs{V-V'}_2\Big)\nonu\\ &\le&2\Big( \abs{E-E'}_1
+\sfrac{2^\de}{2}\;\abs{E-E'}_0^\de
+\sfrac{1}{2}\abs{V-V'}_2^\de+\sfrac{1}{2}\;
\abs{V-V'}_2\Big)\nonu\\ &\le&2\Big(
\abs{E-E'}_1+\abs{E-E'}_0^\de+\abs{V-V'}_2^\de\Big)
\end{eqnarray}
Similarly, by \Ref{dr2} and \Ref{dr3},
\begin{eqnarray}
\abs{e-e'}_2
&\le&
\abs{E-E'}_2+\Qz \abs{\la}\;
 \Big[\abs{e-e'}_1^\de+\abs{e-e'}_2\Big]
\nonu\\
&+&
\Qo\abs{\la}\;(1+G_3)\abs{e-e'}_0 +\Qz \abs{\la}\;
\abs{V-V'}_2
\end{eqnarray}
and
\begin{eqnarray}
\abs{e-e'}_2 &\le&2\Big( \abs{E-E'}_2
+\sfrac{1}{2}\;\abs{e-e'}_1^\de+\sfrac{1}{2}\abs{e-e'}_0
+\sfrac{1}{2}\; \abs{V-V'}_2\Big)\nonu\\ &\le&2\Big( \abs{E-E'}_2
+\sfrac{2^\de}{2}\;\Big[\abs{E-E'}_1^\de+\abs{E-E'}_0^{\de^2}
+\abs{V-V'}_2^{\de^2}\Big]\nonu\\ &&\hspace{2cm}
+\sfrac{1}{2}\Big[2\abs{E-E'}_0+ \abs{V-V'}_2\Big] +\sfrac{1}{2}\;
\abs{V-V'}_2\Big)\nonu\\ &\le&2\Big(
\abs{E-E'}_2+\abs{E-E'}_1^\de+2\abs{E-E'}_0^{\de^2}
+2\abs{V-V'}_2^{\de^2}\Big)
\end{eqnarray}
\end{proof}

\noindent
Part 2 of Theorem \ref{deltasatz} is proven by a multiscale analysis 
in which the function $\KR (E, \la V)$ 
is represented as an infinite series
\begin{equation}
\KR (E, \la V) = \sli_{j<0} \KR_j (E,\la V) ,\end{equation}
where, very roughly speaking, $\KR_j$ is the contribution from
integrating out those fermions that have an energy in the interval
$[M^{j-1},M^j]$. Here $M>1$ and $j<0$, so the limit $j \to \infty$
corresponds to momenta on the Fermi surface.

\begin{theorem}\label{skalensatz}
Let $\dez,\gz,\waz>0$ and $\Gz>\max\{\gz,\waz\}$. Let $\veps>0$
be as in Lemma \ref{radderiv}.
Let $\estart \in\Esset (\dez,\gz,\Gz,\waz)\cap C^3$ and let 
\begin{equation}
\radiEset = \{ e\in\Esset (\dez/2,\gz/2,2\Gz,\waz/2 ): 
\abs{e-\estart}_2 < \veps, \abs{e-\estart}_{3,r} < 1\}.
\end{equation}
There is a $0<\Impeps<1$ such that,
for each $R \in\bN$, there is $\Qt > 0$ ($\Qt$ is uniform on $\Esset$!)
such that for all $e_0,e_1\in \radiEset$ and all $j < 0$, 
\begin{eqnarray}
\label{line1}
\abs{K_j^{(R)}(e)}_{3,r} &=&
\abs{K_j^{(R)}(e)}_2 \le \Qt \abs{\la}  M^{\Impeps j}
\\
\label{line3}
\abs{K_j^{(R)}(e_1)-K_j^{(R)}(e_0)}_1 &\le& \Qt \abs{\la} 
\Big(M^{-1.1\, j} \abs{e_1-e_0}_0+M^{\Impeps j} \abs{e_1-e_0}_1\Big)
\end{eqnarray}
and
\begin{eqnarray}\label{line4}
&& \abs{K_j^{(R)}(e_1)-K_j^{(R)}(e_0)}_{3,r} =
\abs{K_j^{(R)}(e_1)-K_j^{(R)}(e_0)}_{2} \\
&\le&
\Qt \abs{\la}
\left( 
M^{-2.1\,j}\abs{e_1-e_0}_1
 + M^{\Impeps j}\sup\limits_{t \in [0,1]} \abs{e_t}_{3,r} \abs{e_1-e_0}_0
+ M^{\Impeps j} \abs{e_1-e_0}_2 
\right)
\nonu
\end{eqnarray}
Moreover, for all $e\in\Esset (\dez/2,\gz/2,2\Gz,\waz/2)$ and all $V_1,V_2\in \Vset$,
\begin{equation}\label{line5}
\abs{\KR_{j} (e, \la V_1) - \KR_{j} (e, \la V_2)}_2
\le \Qt \abs{\la} M^{\Impeps j}\; 
\abs{V_1-V_2}_2
\end{equation}
\end{theorem}

\noindent 
The proof of Theorem \ref{skalensatz} is given in the next section.
The factors $M^{-1.1 j}$ and $M^{-2.1 j}$ come from bounds  of the
type $M^{-j} |j|^\alpha \le \const\!\!(\alpha)\, M^{-1.1 j}$.

\bigskip
\begin{prooff}
Eq.\ \Ref{line1} implies \Ref{r1} when summed over $j$,
with $\DR = \Qt \sfrac{M^{-\Impeps}}{1-M^{-\Impeps}}$.
Eq.\ \Ref{dr0}
was proven in \ONE\ (Theorem \ONE.3.5). Again by summation, 
\Ref{line5} implies continuity in the interaction $V$.
 
Denote, for brevity, $K_j(e) = \KR_j (e,\la V)$.
To prove \Ref{dr1}, with $\de=\sfrac{\Impeps}{3}$, we split the sum over $j$ in two parts.
If $j$ is such that $|e_1-e_0|_0 \le M^{2j}$, then the 
inequality 
\begin{equation}
|e_1-e_0|_0 \le \big(M^{2j}\big)^{1-\Impeps/3}{|e_1-e_0|_0}^{\Impeps/3}
\end{equation}
implies, by \Ref{line3},
\begin{eqnarray}
\abs{K_j(e_1) - K_j(e_0)}_1 
&\le&\Qt  \abs{\la} \left( M^{(0.9-2\Impeps/3)j}|e_1-e_0|_0^{\Impeps/3} 
+M^{\Impeps j} \abs{e_1-e_0}_1\right)\nonu\\
&\le&\Qt  \abs{\la} \left( M^{0.2\, j}|e_1-e_0|_0^{\Impeps/3} 
+M^{\Impeps j} \abs{e_1-e_0}_1\right)
\end{eqnarray}
and hence
\begin{equation}
\sli_{j\le 0 \atop |e_1-e_0|_0 \le M^{2j}} \;
\abs{K_j(e_1) - K_j(e_0)}_1 \le
\Qtt \abs{\la} \left(\abs{e_1-e_0}_0^{\Impeps/3}+ \abs{e_1-e_0}_1 \right)
\end{equation}
with $\Qtt=\Qt\sfrac{1}{1-M^{-\Impeps'}}$, 
where $\Impeps' = \min\{ 0.2, \Impeps/3\}$.
If $j$ is such that $|e_1-e_0|_0 > M^{2j}$, then 
$|e_1-e_0|_0^{-\Impeps/3} \le M^{-2\Impeps j/3}$  and therefore, by \Ref{line1},
\begin{equation}
\frac{\abs{K_j(e_1) - K_j(e_0)}_1 }{|e_1-e_0|_0^{\Impeps/3}} 
\le 2 M^{-2\Impeps j/3} \max\limits_{p=1,2}\{|K_j (e_p)|_1\}
\le 2\Qt \abs{\la} M^{\Impeps j/3}
\end{equation}
so 
\begin{equation}
\sli_{j \le 0 \atop |e_1-e_0|_0 > M^{2j}} \;
\abs{K_j(e_1) - K_j(e_0)}_1 \le
2\Qtt \abs{\la} \;|e_1-e_0|_0^{\Impeps/3}
.\end{equation}
To prove \Ref{eq35}, with $\de=\sfrac{\Impeps}{4}$, we split the sum over $j$ at $|e_1-e_0|_1 = M^{3j}$.
This time, writing $S_3=\sup\limits_{t \in [0,1]} \abs{e_t}_{3,r}$,
and using
\begin{equation}
|e_1-e_0|_1 \le \big(M^{3j}\big)^{(1-\Impeps/4)}|e_1-e_0|_1^{\Impeps/4}
\end{equation}
when $|e_1-e_0|_1 \le M^{3j}$ gives, by \Ref{line4},
for the $j$ with $|e_1-e_0|_1 \le M^{3j}$, 
\begin{eqnarray}
&&\hspace{-1cm}
\abs{K_j(e_1) - K_j(e_0)}_{3,r} 
\nonu\\
&\le&
\Qt \abs{\la}
\Big( M^{-2.1\,j}\abs{e_1-e_0}_1 + M^{\Impeps j}S_3 \abs{e_1-e_0}_0
+ M^{\Impeps j} \abs{e_1-e_0}_2 \Big)
\nonu\\
&\le&
\Qt \abs{\la}
\Big( 
M^{(0.9-3\Impeps/4)j}|e_1-e_0|_1^{\Impeps/4} 
+ M^{\Impeps j}S_3 \abs{e_1-e_0}_0+ M^{\Impeps j} \abs{e_1-e_0}_2 
\Big)
\nonu\\
&\le&
\Qt \abs{\la}
\Big( 
M^{0.15\,j}|e_1-e_0|_1^{\Impeps/4} 
+ M^{\Impeps j}S_3 \abs{e_1-e_0}_0+ M^{\Impeps j} \abs{e_1-e_0}_2 
\Big)
\end{eqnarray}
so 
\begin{eqnarray}
\sli_{j \le 0 \atop |e_1-e_0|_1 \le M^{3j}}\hspace{-1cm} &&
\abs{K_j(e_1) - K_j(e_0)}_{3,r} \nonu\\
&\le&
\Qf\abs{\la}\; \Big( |e_1-e_0|_1^{\Impeps/4} + S_3 \abs{e_1-e_0}_0+ \abs{e_1-e_0}_2 \Big)
\end{eqnarray}
with $\Qf=\Qt\sfrac{1}{1-M^{-\Impeps'}}$, 
where $\Impeps' = \min\{ 0.15, \Impeps/4\}$.
If $j$ is such that $|e_1-e_0|_1 > M^{3j}$, then 
${|e_1-e_0|_1}^{-\Impeps/4} \le M^{-3\Impeps j/4}$  and therefore, by \Ref{line1},
\begin{equation}
\frac{\abs{K_j(e_1) - K_j(e_0)}_{3,r} }{|e_1-e_0|_1^{\Impeps/4}} 
\le 2 M^{-3\Impeps j/4} \max\limits_{p=1,2}\{|K_j (e_p)|_{3,r}\}
\le 2\Qt \abs{\la} M^{\Impeps j/4}
\end{equation}
so 
\begin{equation}
\sli_{j \le 0 \atop |e_1-e_0|_1 > M^{3j}} \;
\abs{K_j(e_1) - K_j(e_0)}_{3,r} \le
2\Qf \abs{\la} \;|e_1-e_0|_1^{\Impeps/4}
.\end{equation}
\end{prooff}

\newpage

\section{Bounds with scales -- proof of Theorem \ref{skalensatz}}
\label{sect4}

The counterterm is the localization of a selfenergy function,
\begin{equation}
K_j^{(R)} (e,\la V, \Sp) = \ell_e Y_j^{(R)} (e,\la V,p_0,\Sp).
\end{equation}
The renormalized tree expansion gives $Y_j^{(R)}$ explicitly as
\begin{equation}\label{eeexplicit}
Y_j^{(R)} (e,\la V, p) = - \sli_{r=1}^R \la^r \sli_{G} \sli_{T
\sim G} \pli_{f \in T}\frac{1}{n_f!} \sli_{J \in \cJ(T,j,G)}
\Val(G^J)(p)
\end{equation}
where $G$ is summed over all one--particle irreducible (1PI)
Feynman graphs with two external legs and $r$ interaction
vertices. We now briefly describe the genesis of this formula 
as well as the meaning of $T$, $J$, $\cJ(T,j,G)$ and $\Val(G^J)$.
For the details, see, e.g., \cite{FST1}.

The formula is generated by 
 successive applications of renormalization group maps,
as follows (for details, see Section 2.3 of \ONE).
The covariance corresponding to the quadratic part of the action
is expressed as an infinite sum $C=\sum_{j < 0} C_j$, 
where the {\em single--scale covariance}, $ C_j$, is supported 
in the subset of $\bR \times \cB$ where $M^{j-2}\le|\I p_0 - e(\Sp)|\le M^j$
(see Section 2.1 of \ONE). An infrared cutoff $I < 0$ is introduced 
by restricting the sum to $j \ge I$.  Correspondingly, the Gaussian 
integral with the cutoff covariance is expressed as an $|I|$--fold integral
\begin{equation}
\int f(\varphi)\ d\mu_{\Si_{0<j\le I}C_j}(\varphi)
=\int\cdots\int f\big(\smsum_{j=1}^I\varphi_j\big)\ \smprod_{j=1}^I
d\mu_{C_j}(\varphi_j)
\end{equation}
with respect to the Gaussian measures of covariance $C_1,\ \cdots,\ C_I$.
Fields with lower and lower energy scales are integrated out one scale 
after the other. The Gaussian integral with covariance $C_j$ generates 
an {\em effective interaction} on scale $j$. The integral kernels of 
the effective action on scale $j$ are given by a sum of values of Feynman 
graphs whose vertex functions are the integral kernels of the 
effective action on scale $j$ and whose propagators are $C_j$.

The kernel of the part of the effective interaction on scale $j$ that is
quadratic in the fields is renormalized by subtracting from it the part of
the counterterm whose value is $\ell_e$ applied to the kernel. The 
renormalized two--legged kernel is called an $r$--fork
of scale $j$. The remaining part of the counterterm is the sum of all 
$c$--forks of scale $j$. See Section 2.3 of \ONE.

The structure of the iteration is represented by GN (Gallavotti--Nicol\`o)
trees in a natural way. Each graph $G$ contributing to the effective 
interaction at scale $j$ has associated to it a GN tree, $T$. Each fork, 
$f$, in the tree represents a connected subgraph $G_f$ of $G$. The subgraph
was introduced as a vertex contributing to the effective interaction of
some scale $j_f$.  Hence each fork of $T$ carries a label, $j_f$, giving 
its scale and, if $G_f$ is two--legged, a label specifying it as an $r$--fork 
or a $c$--fork. The fork of $T$ corresponding to the entire graph $G$
is called the root of $T$ and its scale, $j$, the root scale of $T$.  The
lines of $T$ give the partial ordering of the forks of $T$ induced by the
partial ordering of subgraphs of $G$ by inclusion. If $\pi(f)$ is the fork
immediately below $f$ in the partial ordering of $T$, then
\begin{eqnarray}\label{scalelimits}
&I\le j_f \le j_{\pi(f)}\qquad&\mbox{ if $\pi(f)$ is a c--fork}
\nonu\\
&1\ge j_f > j_{\pi(f)}\qquad&\mbox{ otherwise}
\end{eqnarray}

The labelling $J$ of $G$ assigns a scale $0<j_l\le I$ to
every line $l$ of $G$ and a scale $0<j_f\le I$ to
every fork $f$ of $T$. The set $\cJ(T,j,G)$ is the set of
labellings determined by the requirements that (a) the root scale is
$j$, (b) (\ref{scalelimits}) is satisfied and (c) if $G_f$ is the smallest
of the subgraphs $G_{f'},\ f'\in T$ that contain the line $l$, then $j_l=j_f$. 

 The value $\Val(G^J)(p)$ of a Feynman graph 
is the integral over momenta of the integrand which is 
a product of propagators associated to the lines and
vertex functions associated to the vertices (see (\ONE.2.54)).
For now, the propagators are given by the covariances 
$C_j$. Later we shall combine strings of two--legged graphs
into single lines, and thereby get more general propagators
on the lines. 

For each $r$, the coefficient of $\la^r$ is a sum of only 
finitely many terms. Thus most perturbative questions
can be reduced to bounding values of individual graphs. 
In some of our estimates in \ONE, however, we also needed to
avoid termwise bounds; this will also play a role in 
this paper.

It was shown in \ONE\ that under general conditions,
the limit
\begin{equation}
K^{(R)}(e,\la V,\Sp) = \lli_{I \to -\infty} \sli_{I \le j < 0} \ell_e
Y_j^{(R)} (e,\la V, \Sp)
\end{equation}
exists and is $C^1$ in $\Sp$ and Fr\` echet differentiable in $e$.

\subsection{Proof of \Ref{line1} and \Ref{line5}}
Eq.\ \Ref{line1} is just a restatement of (\THR.3.110) in 
Theorem \THR.3.11. Because the function $\la_n(j,\varepsilon)$
in (\THR.3.110) is bounded by a constant times a 
power of $|j|$ by Lemma \ONE.2.44 $(v)$,
any $\Impeps < 1/3$ will do. 

To see \Ref{line5}, we note that 
the value of any graph $G$ contributing to $K_j^{(R)}$ in
\Ref{eeexplicit} contains a product of factors $V$ associated to
the vertices. The localization operator $\ell_e$ does not depend
on $V$, and the expression \Ref{eeexplicit} is linear in
$\Val(G)$. Let $G$ be a graph contributing to $K_j^{(R)}$.
By the discrete product rule (\TWO.3.126),
the corresponding graph contributing to the difference on the
left hand side of \Ref{line5} has a difference $V_1-V_2$ 
instead of $V$ in one factor.  Because all that happens to
the vertex functions in the proofs is that they get differentiated
(at most twice), and because the estimate is linear in each
vertex function, \Ref{line5} follows trivially from the proofs in
\ONE--\THR.

\subsection{Weaker hypotheses for the proof of \Ref{dr0}, \Ref{line3}, and \Ref{line4}}

The bounds \Ref{dr0}, \Ref{line3}, and \Ref{line4} hold under much
weaker hypotheses than those stated in Theorem \ref{skalensatz}.
In this section, we prove them under hypotheses that are
only slightly stronger than those of \ONE. 
In particular, we shall need
neither convexity nor symmetry under $\Sp \to -\Sp$
nor the requirement that the Fermi surface be small in the sense that
$S_E \subset \cF_2$. In fact, it need not even be connected.

Let
\begin{description} 
\item\hspace{0.6cm}
$\cN \subset \cB$ be an open set whose boundary has finitely many connected
components, each of which is a $C^\infty$ $(d-1)$--dimensional submanifold of
$\cB$
\item\hspace{0.6cm}
 $u$ be a unit $C^\infty$ vector field on a neighbourhood of
the closure of $\cN$ that is transverse to the boundary of $\cN$
\item\hspace{0.6cm}
 $e_0, e_1 \in C^0(\cB,\bR)\cap C^2(\cN,\bR)$ 
\end{description}
We assume that there are constants $\de_0,\uz,\Cvol,\Impeps > 0$ such that,
for all $s\in [0,1]$, $e_s = (1-s) e_0 + s e_1$ has the following
properties.

\begin{description}

\item[F1]
The set $S_{e_s} = \{ \Sp \in \cB : e_s(\Sp) =0\}$
satisfies $S_{e_s} \subset \cN$ and the distance of $S_{e_s}$ to
$\cB \setminus \cN$ is bounded below by $\de_0$.

\item[F2]
For all $\Sp \in \cN$, 
\begin{equation}\label{cDudef}
\cD_u e_s(\Sp) = u(\Sp) \cdot \nabla e_s(\Sp) > \uz.
\end{equation}

\item[F3]
For $\veps > 0$, let $\Umg(e, \veps) = \{ \Sp \in \cB: |e(\Sp) |
\le \veps\}$. For all $0< \veps_1 \le \veps_2 \le \veps_3$,
and all $\Sq \in \cB$,
\begin{equation}
\ili_{\Umg(e_s,\veps_1)} \dd \Sp_1 \ili_{\Umg(e_s,\veps_2)} \dd
\Sp_2 \True{|e_s(\pm \Sp_1 \pm \Sp_2 + \Sq )| \le \veps_3}
\le
\Cvol \veps_1\veps_2 {\veps_3}^{2 \Impeps }.
\end{equation}

\end{description}

\noin
These hypotheses imply those imposed in \ONE\ (the volume
improvement exponent $\epsilon$ of \ONE\ equals $2 \Impeps$), 
so the results of \ONE\ apply. Moreover, the stronger hypotheses
stated in Section \ref{dispclassect} imply {\bf F1}--{\bf F3} by the following Lemma.

\begin{lemma}\label{Fgood}
Let $B=B_\veps^{(2)} (\estart)$ be the ball of Lemma \ref{radderiv},
$\cN$ be the annulus $\tilde A$ defined in \Ref{Ringerl} and  
$u=\hat r$, the radial vector field of polar coordinates.
Then there are constants $\de_0,\uz,\Cvol,\Impeps > 0$ such that
{\bf F1}--{\bf F3} hold for all $e_0,e_1 \in B$. 
\end{lemma}

\begin{proof}
$B$ is convex, so for all $s \in [0,1]$, 
$e_s=(1-s)e_0 + s e_1 \in B\subset\Esset (\de_0/2,g_0/2,2G_0,\omega_0/2)$.
{\bf F1} is obvious by the definition of $\Esset $.
{\bf F2} follows directly from 
Lemma \ref{radderiv}, with $u_0 = g_1$. 
{\bf F3} follows from Theorem \TWO.1.1 by the usual 
Taylor expansion which is described in 
(\ONE.A.2)--(\ONE.A.6). 
\end{proof}

\noin
{\bf F2} implies that there is $g_0 > 0$ such that for all $s \in
[0,1]$ and all $\Sp \in \cN$,  $|\nabla e_s(\Sp)| > g_0$. For a
fixed $e$, the converse is proven in Lemma \ONE.2.1.

\begin{lemma}\label{Fcoords}
Let $\cN_c$ be a connected component of $\cN$ which has a nonempty intersection
with $S_{e_s}$ for some $0\le s\le 1$. 
\begin{enumerate}
\item
 The boundary of $\cN_c$ has precisely two connected components. 
These two components are diffeomorphic.
\item
Denote by $S$ one of the two components of the boundary of $\cN_c$. 
There is, for each $0\le s\le 1$, a $C^2$ bijection $\Spi_s$ from 
a neighbourhood of $\{0\}\times S$ in $\bR\times S$ to $\cN_c$
such that $e_s (\Spi_s(\rh,\th))=\rh$,
$\sfrac{\del \Spi_s}{\del \rho} (\rh,\th) $ is parallel to 
$u(\Spi_s(\rh,\th))$ and
\begin{equation}
\frac{1}{\sup_{s,\Sp} |\nabla e_s|} \le 
\abs{\frac{\del \Spi_s}{\del \rho} } \le 
\frac{1}{u_0}.
\end{equation}
\end{enumerate}
\end{lemma}
\begin{proof} Denote by $B_1,\ldots, B_n$, the connected components of
the boundary of $\cN$. Since $e_0 \in C^0(\cB,\bR)$ and $\cB$ is compact,
$e_0$ is bounded above and below on $\cN$. By {\bf F2}, the value of $e_0$
changes at a rate of at least $u_0$ per unit time along each trajectory of 
the vector field $u$. Hence each trajectory must start on some $B_i$ and
end on some $B_j$. Because $u$ is transverse to the boundary of $\cN$
and $B_i$ and $B_j$ do not themselves have boundaries,
each trajectory starting on $B_i$ and ending on $B_j$ has an open neighbourhood
\vadjust{\null\hfil\figput{fig4}}
in $\cN$ that is a union of trajectories starting on $B_i$ and ending on $B_j$.
Let, for each $1\le i,j\le n$, $\cN_{i,j}$ be the set of all points of
$\cN$ that lie on a trajectory which starts on $B_i$ and ends on $B_j$.
Then the $\cN_{i,j}$'s are all open and mutually disjoint and their union
is $\cN$. Hence each $\cN_{i,j}$ is either empty or a connected component
of $\cN$. 

We claim that if  $\cN_{i,j}$ has a nonempty intersection with $S_{e_s}$, 
then $i\ne j$. By {\bf F1}, $e_s$ may not vanish in a neighbourhood of the
boundary of $\cN$ and hence must be of uniform sign near each $B_k$.
If $e_s$ has the same sign, say positive, near both $B_i$ and $B_j$ 
(as will certainly be the case if $i=j$) then, as it vanishes somewhere 
in $\cN_{i,j}$, $e_s$ must have a local minimum somewhere in $\cN_{i,j}$.
This violates {\bf F2}.

Suppose that $\cN_c=\cN_{i,j}$. Then $i\ne j$ and the components of the
boundary of $\cN_c$ are $B_i$ and $B_j$. The map which associates to each
$\Sp\in B_i$ the unique point of $B_j$ that is on the same trajectory
as $\Sp$ is a diffeomorphism, so we have completed the proof of part 1.
For each $\Sp\in\cN_c$, denote by $\Theta(\Sp)$ the unique point of $B_i$
that is on the same trajectory of $u$ as $\Sp$. As $B_i$ is a $C^\infty$
manifold, $u$ is transverse to $B_i$ and the trajectories are $C^\infty$
in their dependence on time and initial conditions, $\Theta(\Sp)$ is $C^\infty$.
The map $\Sp\mapsto\big(e_s(\Sp),\Theta(\Sp)\big)$ is defined and $C^2$
on $\cN_c$, injective (as $e_s$ is strictly monotone on each trajectory
and each trajectory hits a different point of $B_i$) onto a neighbourhood
of $\{0\}\times S$ (since $e_s$ is of opposite sign near $B_i$ and 
 $B_j$ it must vanish once on each trajectory). Furthermore the Jacobian
of this map is nonsingular at each $\Sp\in\cN_c$ by {\bf F2} and 
the transversality of $u$ at $B_i$. We may thus take  $\Spi_s$ to be the
inverse of this map.

\end{proof}

\noin
Let $\SP (e_s, \Sp)=\Spi_s\big(0,\Theta(\Sp)\big)$ be the projection on $S_{e_s}$, and let
$\ell_{e_s}$ denote the localization operator for $e_s$, as given
by Definition \ONE.2.6. 
Then $\cD_u \ell_{e_s} = 0$ for all $s \in [0,1]$.
Under the hypotheses of Section \ref{sect1}, and  if $u$ is
chosen to be the radial field $u=\hat r$, 
$\SP$ agrees with the projection $\pp(r,\th) \mapsto \pp(\rF(e_s,\th),\th)$ 
in a neighbourhood of the Fermi surface. 

We now take a fixed $V \in \Vset$ and prove bounds that are 
uniform on $\Vset$.
Thus we again drop the $\la V$ from the notation.

\begin{theorem}\label{oisonacha}
Under the hypotheses F1--F3, there are constants $\QQzer$ and
$\QQone$, depending on $G=\sup_s |e_s|_2$, $\Cvol$, $\Impeps$,
$R$, $r_0$, and $u_0$, such that
\begin{eqnarray}
\abs{\KR (e_1) - \KR(e_0)}_0 &\le& \QQzer \abs{\la}\; \abs{e_1-e_0}_0
\label{nulbo}
\\
\abs{K_j^{(R)}(e_1)-K_j^{(R)}(e_0)}_1 \!\!&\le&\!\! \QQone \abs{\la}
\Big(M^{-1.1\, j} \abs{e_1-e_0}_0+M^{\Impeps j} \abs{e_1-e_0}_1\Big) .
\;
\label{einbo}
\end{eqnarray}
If for all $s \in [0,1]$ the norm $|e_s|_{3,r} = |e_s|_2 +
\norm{\cD_u e_s}_2$ is finite, then
\begin{eqnarray}\label{zwobo}
&& \abs{K_j^{(R)}(e_1)-K_j^{(R)}(e_0)}_{3,r} =
\abs{K_j^{(R)}(e_1)-K_j^{(R)}(e_0)}_{2} \\ &\le& \QQone \abs{\la}
\left( M^{-2.1j}\abs{e_1-e_0}_1
 + M^{\Impeps j}\sup\limits_{t \in [0,1]} \abs{e_t}_{3,r} \abs{e_1-e_0}_0
+ M^{\Impeps j} \abs{e_1-e_0}_2 \right). \nonu
\end{eqnarray}
\end{theorem}

\noin
By Lemma \ref{Fgood}, Theorem \ref{oisonacha} implies 
\Ref{dr0}, \Ref{line3}, and \Ref{line4}, with
$\Qz=\QQzer$ and $\Qt = \QQone$. 

\subsection{Proof of Theorem \ref{oisonacha}}
\label{th4proof}

\bigskip\noin
{\bf Dropping uniform constants in the notation: } We introduce
the notation $A \leco B$ meaning that $A \le \const B$ where the
constant depends only on $G$, $\Cvol$, $\Impeps$, $R$, $r_0$, and
$u_0$ (thus in particular the constant is uniform on $\Esset$).
For instance, we have, for $p \le 3$, $\abs{FG}_p \leco \abs{F}_p
\abs{G}_p$, and $\abs{e}_2 \leco 1$ if $e \in \Esset$.

\bigskip\noin
For a function $F$ that depends on $e$, let $D_h  F$ denote
the directional derivative of $F$ with respect to $e$,
$D_h F = \frac{\del}{\del\al} F(e+\al h)\mid_{\al=0}$.
We proved in \ONE\ that $K$ is Fr\' echet differentiable 
in $e$, so these derivatives exist. Moreover, Fr\' echet
differentiability holds for all quantities in which there
is an infrared cutoff. 

\subsubsection*{Proof of \Ref{nulbo}}
By  \Ref{eeexplicit}, for any $s\in[0,1]$,
\begin{eqnarray}
&& \abs{D_h \left(\ell_{e_s} \sli_{I \le j < 0} Y_j^{(R)} (e_s)\right)}_0
\nonu\\ &\le& \sli_{r=1}^R |\la|^r \sli_{G} \abs{ \sli_{I \le j <
0} \sli_{T \sim G} \pli_{f \in T}\frac{1}{n_f!} \sli_{J \in
\cJ(T,j,G)} D_h \left(\ell_{e_s} \Val(G^J)(e_s)\right)}_0
\end{eqnarray}
By (\ONE.3.35), there is a constant, depending only on $G$ and on
the constants given in the Lemma, such that
\begin{equation}
\abs{D_h \left(\ell_{e_s} \sli_{I \le j < 0} Y_j^{(R)} (e_s)\right)}_0 \le
\sli_{r=1}^R |\la|^r \sli_{G} \const (G) |h|_0.
\end{equation}
For fixed $R$, the sum over graphs $G$ contains finitely many
terms, so
\begin{equation}\label{summed}
\abs{D_{e_1-e_0} \left(\ell_{e_s} \sli_{I \le j < 0} Y_j^{(R)} (e_s)\right)}_0
\leco |\la| \; |e_1-e_0|_0
\end{equation}
uniformly in $I$ and $s$. Thus \Ref{nulbo} follows by
\begin{eqnarray}
\sli_{j < 0} (K_j^{(R)} (e_1) - K_j^{(R)} (e_0) ) &=& \ili_0^1 \dd
s\; \frac{\del}{\del s} \ell_{e_s} \sli_{j < 0} Y_j^{(R)} (e_s)
\nonu\\ &=& \ili_0^1 \dd s\; D_{e_1-e_0} \ell_{e_s} \sli_{ j < 0}
Y_j^{(R)} (e_s) .
\end{eqnarray}

\subsubsection*{Preliminaries for the proof of \Ref{einbo} and \Ref{zwobo}}
To prove the single--scale bounds \Ref{einbo} and \Ref{zwobo}, we
show that for $k \le 2$, the seimnorms 
$\Vert K_j^{(R)} (e_1) - K_j^{(R)} (e_0) \Vert_k$ 
obey bounds with the same right hand side as in
\Ref{einbo} and \Ref{zwobo}. Note that even the bound 
for $k=0$ does not follow from \Ref{summed} because we are now
considering a fixed scale $j$, not a sum over scales, and the
summation over scales provided a cancellation that was important
in the proof of Theorem \ONE.3.5. However, 
the proof does not require very detailed estimates because
the coefficient of
$|e_1-e_0|_{k-1}$ is (up to factors $|j|$, which we bound
by $M^{-0.1j}$)
a factor $M^{-kj}$ larger than the
undifferentiated power counting behaviour $M^j$
of a single--scale
selfenergy contribution like $Y_j^{(R)}$.
This is naive power counting behaviour.
The estimates will
again follow by applying bounds already proven in \ONE.

We now interpolate the difference of the two $K$ functions. The
derivative of $\ell_e$ with respect to $e$ was calculated in Lemma
\ONE.3.1. The interpolation gives
\begin{equation}
K_j^{(R)} (e_1) - K_j^{(R)} (e_0)
=
\ili_0^1 \dd s\; \left( \cY_1(s) -\cY_2(s) \right)
\end{equation}
with
\begin{eqnarray}
\cY_1(s) &=& \ell_{e_s} \left( D_{e_1-e_0}  Y_j^{(R)} (e_s)
\right)
\\
\cY_2(s) &=& \ell_{e_s} \left[ (e_1-e_0) \frac{1}{\cD_u e_s} \cD_u
Y_j^{(R)} (e_s)\right] ,
\end{eqnarray}
with $\cD_u$ defined in \Ref{cDudef}.
Because $K_j$ is the localization of $Y_j$, $\cD_u K_j =0$, so the
first equality in \Ref{zwobo} holds. Thus we have to bound
$\norm{\cY_i}_k$ for $k\in\{0,1,2\}$. 
In the following, we drop the superscript $R$ from $Y_j^{(R)}$. 

\subsubsection*{Estimates for $\norm{\cY_2}_k$}
Let $k=0$. The bound $|\cY_2|_0 \le \abs{e_1-e_0}_0 \sfrac{1}{\uz}
|\cD_u Y_j|_0$ and Theorem \ONE.2.46 (i) imply that
\begin{equation}\label{0bo}
\abs{\cY_2}_0 \leco |j|^R M^{2 \Impeps j} \abs{e_1-e_0}_0 \leco M^{\Impeps
j} \abs{e_1-e_0}_0.
\end{equation}
Let $k=1$. Because
\begin{equation}
\sfrac{\del}{\del p_\al} (\ell_{e_s} F)(p) = \sfrac{\del}{\del
p_\al} F(0,\SP(e_s,\Sp)) = \sli_\be \sfrac{\del \SP_\be}{\del
p_\al}(e_s,\Sp) \left[ \sfrac{\del}{\del q_\be} F(0,\Sq)
\right]_{\Sq=\SP(e_s,\Sp)},
\end{equation}
we have
\begin{equation}
\sfrac{\del}{\del p_\al} \cY_2 (s)(p) = \sli_\be \sfrac{\del
\SP_\be}{\del p_\al}(e_s,\Sp) \; \cX_\be (\SP(e_s,\Sp))
\end{equation}
with
\begin{equation}
\cX_\be (\Sq) = \sfrac{\del}{\del q_\be} \left[ (e_1-e_0)(\Sq) \,
\sfrac{1}{\cD_u e_s (\Sq)} \cD_u Y_j(0,\Sq) \right] .
\end{equation}
Thus
\begin{eqnarray}\label{1bo}
\norm{\cY_2(s)}_1 & \le & d \; \norm{\SP(e_s)}_1 \left(
\norm{e_1-e_0}_1 \sfrac{1}{\uz} \abs{\cD_u Y_j}_0 \right. \nonu\\
&& + \qquad \abs{e_1-e_0}_0 \sfrac{1}{\uz^2} \norm{\cD_u e_s}_1 \abs{\cD_u
Y_j}_0 \nonu\\ && + \qquad \left. \abs{e_1-e_0}_0 \sfrac{1}{\uz}
\norm{\cD_u Y_j}_1 \right) \nonu\\ &\leco& \abs{e_1-e_0}_1
\abs{\cD_u Y_j}_0 + \abs{e_1-e_0}_0 \norm{\cD_u Y_j}_1
\end{eqnarray}
because $\norm{\SP (e_s)}_1 \leco 1$ and 
$\norm{\cD_u e_s}_1 \leco \abs{e_s}_2 \leco 1$. 

\bigskip\noin
Let $k=2$.
Because
\begin{eqnarray}
\sfrac{\del^2}{\del p_\ga \del p_\al} \cY_2(p) &=& \sli_\be \left[
\cX_\be (\SP(e_s,\Sp)) \sfrac{\del^2\SP_\be}{\del \Sp_\ga
\del\Sp_\al}(e_s,\Sp)\right. \nonu\\ &+& \left. \sli_\rh (\del_\rh
\cX_\be)(\SP(e_s,\Sp)) \del_\ga \SP_\rh (e_s,\Sp) \del_\al \SP_\be
(e_s,\Sp) \right] ,
\end{eqnarray}
we have
\begin{equation}
\norm{\cY_2(s)}_2 \le d\; \norm{\SP(e_s)}_2 \max\limits_\be
\abs{\cX_\be}_0 + {\norm{\SP(e_s)}_1}^2 \sli_{\be,\rh}
\abs{\del_\rh\cX_\be}_0.
\end{equation}
Because $\norm{\SP(e_s)}_2 \leco 1$ and
\begin{eqnarray}
\abs{\sfrac{\del}{\del q_\rh} \cX_\be (\Sq)} &\leco&
\abs{e_1-e_0}_2 \abs{\cD_u Y_j}_0 + \abs{e_1-e_0}_0 \abs{\cD_u
Y_j}_0 \norm{\cD_u e_s}_2 \nonu\\ &+& \abs{e_1-e_0}_1 \norm{\cD_u
Y_j}_1 + \abs{e_1-e_0}_0 \norm{\cD_u Y_j}_2 ,
\end{eqnarray}
\intremark{
In more detail:
\begin{equation}
\del_\rh \cX_\be (\Sq) = \sfrac{\del^2}{\del q_\be \del q_\rh}
\left( (e_1-e_0)(\Sq) \sfrac{1}{\cD_u e_s(\Sq)} \cD_u Y_j (0,\Sq)
\right) ,
\end{equation}
and so
\begin{eqnarray}
\abs{\sfrac{\del}{\del q_\rh} \cX_\be (q)} &\lenuco&
\norm{e_1-e_0}_2 \sfrac{1}{\uz} \abs{\cD_u Y_j}_0 \nonu\\ &+&
\abs{e_1-e_0}_0 \left( \sfrac{1}{\uz^3} {\norm{\cD_u e_s}_1}^2 +
\sfrac{1}{\uz^2} \norm{\cD_u e_s}_2 \right) \abs{\cD_u Y_j}_0
\nonu\\ &+& \abs{e_1-e_0}_0 \sfrac{1}{\uz} \norm{\cD_u Y_j}_2
\nonu\\ &+& \norm{e_1-e_0}_1 \sfrac{1}{\uz^2} \norm{\cD_u e_s}_1
\abs{\cD_u Y_j}_0 \nonu\\ &+& \norm{e_1-e_0}_1 \sfrac{1}{\uz}
\norm{\cD_u Y_j}_1 \nonu\\ &+& \abs{e_1-e_0}_0 \sfrac{1}{\uz^2}
\norm{\cD_u e_s}_1 \norm{\cD_u Y_j}_1 \nonu\\ && \nonu\\ &\leco&
\abs{e_1-e_0}_2 \abs{\cD_u Y_j}_0 + \abs{e_1-e_0}_0 \abs{\cD_u
Y_j}_0 \norm{\cD_u e_s}_2 \nonu\\ &+& \abs{e_1-e_0}_1 \norm{\cD_u
Y_j}_1 + \abs{e_1-e_0}_0 \norm{\cD_u Y_j}_2 \nonu\\ && \nonu\\
&\leco& \abs{e_1-e_0}_2 \norm{Y_j}_1 + \abs{e_1-e_0}_0
\norm{Y_j}_1 \norm{\cD_u e_s}_2 \nonu\\ &+& \abs{e_1-e_0}_1
\norm{Y_j}_2 + \abs{e_1-e_0}_0 \norm{\cD_u Y_j}_2
\end{eqnarray}
}
we have
\begin{eqnarray}\label{2bo}
\norm{\cY_2(s)}_2 &\leco& \abs{e_1-e_0}_2 \, \abs{\cD_u Y_j}_0 +
\abs{e_1-e_0}_1\,  \norm{\cD_u Y_j}_1 \nonu\\ &+& \abs{e_1-e_0}_0
\, \left( \abs{\cD_u Y_j}_0 \, \norm{\cD_u e_s}_2 +  \norm{\cD_u
Y_j}_2 \right) .
\end{eqnarray}
The term $\norm{\cD_u e_s}_2$ is the reason why we have to deal
with functions that have bounded radial derivatives. Because it
arises only from the derivative of the localization operator, it
has got nothing to do with the scale dependence of $Y_j$. 

By \Ref{eeexplicit}, it suffices to bound the contribution from every
1PI two--legged graph $G$ separately. That is, we may replace
$Y_j$ by $W=\sum_{J \in \cJ(T,j,g)} \Val (G^J) $ in \Ref{0bo},
\Ref{1bo}, and \Ref{2bo} if we take a maximum over $G$ and $T$
and multiply by the number of graphs and the number of possible 
$T$'s. 
By Theorem \ONE.2.46 (i),
and using $\la_n(j,\Impeps) M^{\Impeps j} \leco 1$, we have
\begin{eqnarray}
\norm{\cY_2(s)}_1 &\leco& \abs{e_1-e_0}_1 \, M^{\Impeps j} +
\abs{e_1-e_0}_0\,  M^{(\Impeps-1) j}
\\
\norm{\cY_2(s)}_2 &\leco& \abs{e_1-e_0}_2 \, M^{\Impeps j} +
\abs{e_1-e_0}_1\,  M^{(\Impeps-1) j} \nonu\\ &+& \abs{e_1-e_0}_0 \, \left(
M^{\Impeps j} \, \norm{\cD_u e_s}_2 +  \norm{\cD_u Y_j}_2 \right) .
\end{eqnarray}
Thus $\sup_s \norm{\cY_2(s)}_k$ obey bounds that imply \Ref{einbo}
and \Ref{zwobo} if we can prove that
\begin{equation}
\norm{\cD_u Y_j}_2 \leco M^{-2.1\, j}
\end{equation}
and that 
\begin{equation}
\norm{\cY_1(s)}_1 \leco M^{-1.1 j} \abs{e_1-e_0}_0,
\quad
\norm{\cY_1(s)}_2 \leco M^{-2.1 j} \abs{e_1-e_0}_0.
\end{equation}
To do this, we need to exhibit the structure of the graphs $G$ 
that contribute to $Y_j$ in a little bit more detail.

\subsubsection*{Graphical tools}
Let $G$ be a graph contributing to \Ref{eeexplicit}, $T$ a rooted
tree compatible to $G$, with an $r$ and $c$ labelling assigned to
the forks, and $\cJ(T,j,G)$ the set of labellings of $G$
compatible with $T$ and root scale $j$. Let $\ph$ be the root of
$T$. To every fork $f \in T$ there corresponds a connected
subgraph $G_f$ of $G$, which is a proper subgraph of $G$ for $f >
\ph$. We call $f$ an $m$--legged fork if $G_f$ has $m$ external
legs. In the following we construct a graph $\Ga$, a tree $T'$
compatible with $\Ga$, and a set of labellings $\cJ'$ with the
following properties.

\begin{itemize}

\item
$\Ga$ is two--legged and 1PI, and $\Ga$ has only four--legged
vertices with vertex functions $\hat v$.

\item
The associated tree $T'$ has no 2--legged forks.

\item
The scale assignments in $\cJ'$ are $j_f > j_{\pi(f)}$ for all
$f\in T'$. With propagators associated to $\Ga$ in the way given
below,
\begin{equation}\label{s1}
\sli_{J \in \cJ(T,j,G)} \Val (G^J) = \sli_{J' \in \cJ'(T',j,\Ga)}
\Val (\Ga^{J'})
\end{equation}
Summation over the trees gives
\begin{equation}\label{s2}
\sli_{T \sim G}
\pli_{f\in T} \frac{1}{n_f!} 
\sli_{J \in \cJ(T,j,G)} \Val (G^J) 
= 
\sli_{T' \sim \Ga} 
\pli_{f'\in T'} \frac{1}{n_{f'}!} 
\sli_{J' \in \cJ'(T',j,\Ga)} \Val (\Ga^{J'}) .
\end{equation}

\end{itemize}

\noin
This construction is similar to that of Remark \ONE.2.45,
only simpler, because here we do not aim at tight bounds for the
powers of $|j|$ generated by scale sums of four--legged
subdiagrams.

If no $f > \ph$ is two--legged, then $\Ga=G$, $T'=T$, $\cJ'=\cJ$.
Otherwise, let $f_1, \ldots f_n > \ph$ be all minimal two--legged
forks of $T$. That is, there is no two--legged fork $f'$ with $\ph
< f' < f_i$. Let $\tilde T$ be the tree where the subtrees $T_i$
rooted at the forks $f_i$ are replaced by leaves $\la_i$. To
obtain the corresponding graph $\tilde G$, replace $G_{f_i}$ by a
two--legged vertex $v_i$ with ($j_{\pi(f_i)}$--dependent)
vertex function
\begin{equation}\label{2vf}
A_i = \cP_i \sli_{j_{i}} \sli_{J_i \in
\cJ(T_i,j_{\pi(f_i)},G_{f_i})} \Val (G_{f_i}^{J_i}).
\end{equation}
The projection $\cP_i$ is $\ell_{e_s}$ if $f_i$ is a $c$--fork and
$1-\ell_{e_s}$ if $F_i$ is an $r$--fork of $T$. The summation
range is $j_i > j_{\pi(f_i)}$ if $f_i$ is an $r$--fork and $j_i
\le j_{\pi(f_i)}$ if $f_i$ is a $c$--fork.

Because all $c$--forks have now been replaced by vertices (or
hidden inside two--legged vertices), $\tilde\cJ = \{
J\vert_{\tilde T}: J \in \cJ(T,j,G)\}$ consists only of labellings
with $j_f > j_{\pi(f)} $ for all $f \in \tilde T$. With the
standard definition of the value of a labelled graph (see, e.g.,
(\ONE.2.54)),
\begin{equation}
\sli_{J \in \cJ(T,j,G)} \Val(G^J) = \sli_{\tilde J \in \tilde
\cJ(\tilde T, j, \tilde G)} \Val\left({\tilde G}^{\tilde
J}\right).
\end{equation}
The graph $\tilde G $ is not yet what we want because the graph
$G_{f_i}$ whose value appears in \Ref{2vf} is not necessarily 1PI
and because $\tilde G$ may contain two--legged vertices. 
In order to apply Theorem \ONE.2.46, we want to reduce all vertex
functions of two--legged vertices to sums over values of 1PI
graphs.

If $f_i$ is a $c$--fork, $G_{f_i}$ is 1PI because otherwise
$\ell_{e_s}$ of its value would vanish. If $f_i$ is an $r$--fork,
$G_{f_i}$ may be 1PR; then $\cP_i \Val (G_{f_i}^J) = \Val (G_{f_i}^J)$
and it is a string of two--legged subgraphs, 
some of which may be single--scale insertions (SSI's)
defined in Remark \ONE.2.45.
Momentum conservation, the scale structure on $T$, and the support
properties of the cutoff function fix the scale of the lines
connecting the 1PI pieces to $j_{\pi(f_i)}+1$. 
When every $r$--fork corresponding to an 1PR graph is replaced
by its string as above, the only changes to $\tilde G$ 
are that additional two--legged vertices may appear 
and that, besides the cases $\cP_i=\ell_{e_s}, 1 -\ell_{e_s}$, 
there is the third
case $\cP_i=1$ for SSI's, with the scale sum for a SSI consisting
only of the one term where all scales are $j_{\pi(f_i)}+1$ (see
Remark \ONE.2.45 for details).

Let $\Ga$ be the graph where all strings of two--legged subgraphs
are replaced by single lines, and $T'$ be the tree in which all
leaves of $\tilde T$ that correspond to two--legged vertices of
$\tilde G$ are removed. For a line $\ell$ of $\Ga$, let $j_\ell$
be the minimum over all $j_{\tilde \ell}$, where $\tilde \ell$ runs
over the lines of $\tilde G$ on the string $\si_\ell$ in $\tilde G$
replaced by $\ell$. The propagator associated to $\ell$ is
\begin{equation}\label{Gaprop}
  S_{\ell,j_\ell} (p) =
  \sli_{(j_{\tilde \ell})_{\tilde \ell\;{\rm on}\;\si_\ell}}
  \pli_{\tilde \ell \;{\rm on}\; \si_\ell} C_{j_{\tilde \ell}} (p)
  \pli_{v \;{\rm on}\; \si_\ell} A_v
\end{equation}
where the summation is over all scale assignments $j_{\tilde \ell}
\in \{j_\ell,j_\ell+1\}$ that are compatible with $\tilde T$, and,
if $n$ propagators appear in the product, $n-1$ factors $A_v$
appear. By construction, \Ref{s1} and \Ref{s2} hold.

\begin{lemma}\label{strider}
Let $\al$ be a multiindex with $w=|\al| \le 1$. Then the
propagators $ S_{\ell,j_\ell}$ given by \Ref{Gaprop} satisfy
\begin{equation}\label{Gapropbou}
  \abs{D^\al  S_{\ell,j_\ell} (p) } \leco
  M^{-j_\ell(1+w) + j_\ell \Impeps g}
  \True{|\I p_0 - e_s(\Sp)| \le M^{j_\ell}} 
\end{equation}
where $g$ is the number of $c$--forks plus the number of SSI on
the string $\si_\ell$ corresponding to $S_{\ell,j_\ell}$.
\end{lemma}

\begin{proof}
The support condition follows directly from that of $C_{j_\ell}$.
We now bound the functions $A_v$ and their first derivatives. This
is a direct application of Theorem \ONE.2.46 (i), which states
(with $\veps = 2 \Impeps$) 
that if $G$ is two--legged and 1PI, then for all $r \in \{
0,1,2\}$,
\begin{equation}\label{I.2.46.i}
  \sli_{J \in \cJ(T,j,G)} \abs{\Val G^J}_r \leco |j|^{n_G} M^{j
  (1+2\Impeps -r)} \leco M^{j (1+\Impeps-r)} .
\end{equation}
Let $w \in \{0,1\}$ and $\al $ be a multiindex with $|\al|=w$. For
$v$ corresponding to an $r$--fork and for $p$ such that $\abs{\I
p_0 - e_s(\Sp)} \le M^{j_\ell}$,
\begin{equation}\label{rfork}
\abs{D^\al A_v (p)} \leco \sli_{j >j_{\ell}} \abs{ \sli_{J \in \cJ
(T_i,j,G_{f_i})} D^\al (1- \ell_{e_s}) \Val (G_{f_i}^J)(p)}
\end{equation}
For $w=0$, Taylor expansion gives the renormalization gain $M^{j_\ell}$ and 
one derivative acting on $\Val (G_{f_i}^J)$. By
\Ref{I.2.46.i}, with $r=1+w=1$,
\begin{equation}\label{e67}
  \abs{D^\al A_v (p)} \leco M^{j_\ell} \sli_{j > j_\ell}
  M^{j(1+\Impeps-1)}
  \leco   M^{j_\ell}.
\end{equation}
For $w=1$, we estimate the $1$ and $\ell_{e_s}$ terms separately. 
By \Ref{I.2.46.i},
\begin{equation}
  \abs{D^\al A_v (p)} \leco 2 \sli_{j > j_\ell}
  M^{j(1+\Impeps-1)}
  \leco   1.
\end{equation}
For $v$ corresponding to a $c$--fork,
\begin{equation}\label{cfork}
  \abs{A_v}_w \leco \sli_{j \le j_\ell}
  \abs{ \sli_{J \in \cJ(T_i,j,G_{f_i})}
  \ell_{e_s} \Val (G_{f_i}^J)}_w ,
\end{equation}
so \Ref{I.2.46.i} implies
\begin{equation}\label{e70}
  \abs{A_v}_w \leco \sli_{j \le j_\ell} M^{j(1+\Impeps-w)} \leco
  M^{j_\ell(1+\Impeps-w)}.
\end{equation}
The estimate for $v$ corresponding to an SSI is similar to that of
a $c$--fork, except that there is not even a scale sum to do
because the scales are all fixed in an SSI. Using the product rule
for derivatives acting on \Ref{Gaprop} and using that
\begin{equation}\label{standardCj}
  \abs{D^\al C_j(p)} \leco M^{-j(1+|\al|)} \True{|\I p_0 -e_s(\Sp)| \le M^j}
\end{equation}
we get the statement of the Lemma.
\end{proof}

\noin
Lemma \ref{strider} gives us control over first order derivatives
of the propagators $S_{\ell,j_\ell}$ with respect to momentum. 
The next lemma will imply that we can always arrange the integral
for the value of a graph contributing to $Y_j$ such that every line of the graph gets
differentiated at most once, even if we take three derivatives 
with respect to the external momentum. 

In \ONE, Definition 2.19, we introduced the notion of 
overlapping graphs. A graph is overlapping if there is a line 
$\ell$ of $G$ which is part of two independent 
(non self--intersecting) loops. 
We say that the line $\ell$ is part of the two overlapping loops.

\begin{lemma}\label{graphlem}
Let $G$ be a two--legged 1PI graph with two external vertices
$v_1$ and $v_2$. Let all vertices of $G$ have an even incidence
number. Let $T$ be any spanning tree of $G$, and let $\th$ be the
linear subtree of $T$ corresponding to the unique path from $v_1$
to $v_2$ over lines of $T$. Then every line $\ell \in \th$ is part
of two overlapping loops generated by lines $\ell_1\not\in T$ and
$\ell_2\not\in T$. 
For $i \in \{1,2\}$, the graph $T_i$, obtained from $T$ by removing
$\ell$ and adding $\ell_i$, is a spanning tree for $G$.
\end{lemma}

\begin{proof}
Let $\ell$ be a line of $\th$. Cut $\ell$ to get a four--legged
graph $F=G-\ell$. Because $G$ is 1PI, $F$ is connected, so there
is a (nonself{}intersecting) path $\pi$ in $F$ that joins the
endpoints of $\ell$. Because $T$ is a tree, $T-\ell$ has two 
connected components, $T_1$ and $T_2$. 
As $T_1 \cup T_2 \cup \pi$ is connected, 
one of the lines on $\pi$, say $\ell_1$, joins $T_1$ and $T_2$,
but is not in $T$. 
Thus $\ell $ is on the loop generated by $\ell_1$. Go back to $G$ and
cut $\ell_1$. The result is a four--legged graph $F' = G -
\ell_1$. Because $\ell_1 \not\in T$, $T$ is still a spanning tree
for $F'$. Cutting $\ell$ does not disconnect $F'$
because if it did, each of the connected components would have to have
three external lines -- one of $G$'s original external lines, 
one end of $\ell_1$ and one end of $\ell$ (as all vertices of $G$ have even
incidence number, all connected graphs must have an even number of external
lines). 
Let $\ell_2$ be a line on the shortest path in $F' - \ell$ connecting 
the endpoints of $\ell$ with $\ell_2$ joining $T_1$ and $T_2$ but 
not in $T$. 
Then $\ell$ is in the loop generated by $\ell_2$. Thus the loops
generated by $\ell_1$ and $\ell_2$ overlap on $\ell$.
\end{proof}

\noin 
It would not have been a loss of generality to
assume that $G$ has no proper two--legged subgraphs. 
In that case, Remark \ONE.2.23 implies that $F'$ is also 1PI.
If $T$ is chosen such that $\th$ is a shortest path from
$v_1$ to $v_2$ in $G$, the statement of the Lemma is an obvious
consequence of Lemma \THR.2.5 (see Figures \THR.2.3--\THR.2.6;
note that the lines from $v_r$ to $v_{r+1}$ and from $v_s$ to
$v_{s+1}$ can be any pair of lines on $\th$). 

\subsubsection*{The bound for $\norm{\cD_u Y_j}_2$}
Because $\norm{\cD_u Y_j}_2 \le \abs{Y_j}_3$, 
it suffices to prove that 
\begin{equation}
\abs{Y_j}_3 \leco M^{-2.1j}.
\end{equation}
By \Ref{eeexplicit}, it suffices to prove the same bound for
\begin{equation}
\cW = \sli_{J \in \cJ(T,j,G)} \Val(G^J).
\end{equation}
All graphs that contribute are two--legged and 1PI, so by \Ref{I.2.46.i}, 
\begin{equation}
\abs{\cW}_2 \leco M^{j(1+2\Impeps-2)} (1+|j|^R) \leco M^{-j} M^{\Impeps j}
\leco M^{-2j},
\end{equation}
so it suffices to bound $\norm{\cW}_3$. Let $\Ga$ be the graph 
associated to $G$ with the properties \Ref{s1} and \Ref{s2}, then
\begin{equation}\label{e123}
\norm{\cW}_3 \le \sli_{J \in \cJ(T',j,\Ga)}
\norm{\Val(\Ga^J)}_3. 
\end{equation}
Let $T$ be a spanning tree for $\Ga$. 
The only factors in the integrand for $\Val \Ga^J$ that can depend on 
the external momentum $q$ are 

\begin{itemize}

\item
vertex functions $\hat v$; the dependence is of the form $\hat v(q-p)$
where $p$ is a loop momentum or a sum of loop momenta 
because $G$ is 1PI and two--legged
(it can happen that $\hat v$ does not depend on any loop momentum;
this is, however, only the case for tadpoles, in which case only $\hat v(0)$
appears). 

\item
propagators $S_{\ell, j_\ell}$ for those $\ell $ that are in the path
on $T$ connecting the external vertices (if there is only one external 
vertex, no propagator depends on $q$). 

\end{itemize}

\noin
We now take three derivatives of $\Val(\Ga^J)$ and use the above lemmas
to avoid having two derivatives acting on any propagator and three 
on any vertex function, as follows.

If $\Ga$ has only one external vertex and is not a tadpole,
we first route $q$ through the $\hat v$ of the external vertex.
We let two derivatives act and then change variables from $p$
to $q-p$ in the loop integral in which $\hat v(q-p)$ appears. 
The third derivative can then not act on this vertex function any more. 
It can act on another vertex function or on a propagator. 

If $\Ga$ has two external vertices, there are two cases, 
depending on where the first derivative acted. 

\begin{enumerate}

\item
The first derivative acts on a vertex function. Take another derivative.
If it acts on the same vertex function, change variables from $p$
to $q-p$ in the loop integral in which $\hat v(q-p)$ appears. 
The third derivative can then not act on this vertex function any more. 
If the second derivative acts on the propagator $S_{\ell, j_\ell}$, 
we change the spanning tree using Lemma \ref{graphlem}. 
The third derivative can then not act on $S_{\ell, j_\ell}$ any more.

\item
The first derivative acts on the propagator $S_{\ell, j_\ell}$. 
We change the spanning tree to $T_1$ by replacing $\ell$ with another 
line $\ell_1$ (this is possible by Lemma \ref{graphlem})
and take another derivative. 
It can act on a propagator on a line $\ell'$ on the path in $T_1$ 
that connects the external vertices ($\ell'=\ell_1$ is possible).
By Lemma \ref{graphlem}, there are two lines, $\ell'_1$ and $\ell'_2$,
such that for $i\in\{1,2\}$, $T'_i$, obtained by replacing $\ell'$ 
by $\ell'_i$ in $T_1$, is still a spanning tree for $\Ga$. 
At most one of $\ell'_1$ and $\ell'_2$ may be $\ell$, 
so we may change to a spanning tree that contains 
neither $\ell$ nor $\ell'$. Once this is done, the third derivative
cannot act on the propagators associated to the lines $\ell$ and $\ell'$.

\end{enumerate}

In summary, the net effect of taking three derivatives in the 
way just described is, by Lemma \ref{strider}, at most a factor
$M^{-3j}$, as compared to standard power counting
(a factor $M^{-3j}$ arises only if all three derivatives
act on propagators; when vertex functions get differentiated,
no factor $M^{-j}$ is produced). 
Because the GN tree $T'$ associated to $\Ga$ has no 2--legged forks, 
the scale sum converges by standard arguments (see Lemma \ONE.2.4
and Remark \ONE.2.5), and is bounded by $|j|^R M^{j}$. Thus
\begin{equation}\label{e124}
\norm{\cW}_3 \leco |j|^R M^{j} M^{-3j} \leco M^{-2.1j}.
\end{equation}

\subsubsection*{The bound for $\norm{\cY_1(s)}_2$}
In the following bounds we keep the tree sums inside of the 
norms. By \Ref{s2}, we thus need to estimate
\begin{equation}
\norm{\sli_{T' \sim \Ga} 
\pli_{f \in T'} \frac{1}{n_f!}
\sli_{J \in \cJ(T',j,\Ga')} 
D_h \Val ( \Ga^{J}) }_k
\end{equation}
for $k \in \{0,1,2\}$, with $h = e_1-e_0$. 
By construction of $\Ga$, $D_h$ acts only on the propagators
$S_{\ell,j_\ell}$. 

\begin{lemma}
For all $s \in [0,1]$ and all lines $\ell$ of $\Ga$
\begin{equation}
\abs{D_h S_{\ell,j_\ell} (p)} \leco
|h|_0\; M^{-2j_\ell} \True{ |ip_0 - e_s (\Sp) | \le M^{j_\ell}}.
\end{equation}
\end{lemma}

\begin{proof}
By definition \Ref{Gaprop}, $D_h$ can act on factors (a) $C_j$, 
(b) $A_v$ coming from an $r$--fork, (c) $A_v$ coming from a $c$--fork,
(d) $A_v$ coming from an SSI.
In the last three cases, by \Ref{2vf}, we have to estimate the norms of
\begin{equation}
\tilde\cW_{i} = 
\cP_i \sli_{j_{i}} 
\sli_{T_i \sim G_{f_i}} 
\pli_{f \in T_i} \frac{1}{n_f!}
\sli_{J_i \in \cJ(T_i,j_{\pi(f_i)},G_{f_i})} 
\Val (G_{f_i}^{J_i}).
\end{equation}
\begin{description}

\item[(a)]
by (\ONE.3.44), 
\begin{equation}
\abs{D_h C_{j_{\tilde\ell}}(p)} \leco 
|h|_0\; M^{-2j_\ell} \True{ |ip_0 - e_s (\Sp) | \le M^{j_{\tilde\ell}}}.
\end{equation}

\item[(b)]
By Lemma \ONE.3.1, 
\begin{equation}
D_h (\ell_{e_s} \cW_i) (e_s)  = \ell_{e_s} (D_h \cW_i) 
- \ell_{e_s} \left(\sfrac{h}{\cD_u e_s} \cD_u \cW_i \right) (e_s)
\end{equation}
so 
\begin{equation}\label{hach}
D_h (1-\ell_{e_s})\cW_i = (1-\ell_{e_s}) D_h \cW_i + 
\ell_{e_s} \left(\sfrac{h}{\cD_u e_s} \cD_u \cW_i \right).
\end{equation}
If $p$ is such that $|ip_0 - e_s (\Sp) | \le M^{j_\ell}$, then 
by Taylor expansion
\begin{equation}
\abs{(1-\ell_{e_s}) D_h \cW_i (p)} \leco M^{j_\ell} \abs{D_h  \cW_i}_1.
\end{equation}
By (\ONE.3.42), this is 
\begin{equation}
\leco \abs{h}_0\; M^{j_\ell} \sli_{j > j_\ell} 
M^{j(2\Impeps -1)} |{j_\ell}|^{R_i}
\leco \abs{h}_0 |{j_\ell}|^{R_i} M^{2\Impeps j_\ell} 
\leco \abs{h}_0
\end{equation}
with $R_i$ the number of vertices of $G_{f_i}$. 
The second term in \Ref{hach} is bounded by 
\begin{equation}
\abs{\sfrac{h}{\cD_u e_s} \cD_u \cW_i}_0 \leco \abs{h}_0 \abs{\cD_u \cW_i}_0
\leco  \abs{h}_0 \abs{\cW_i}_1 \leco  \abs{h}_0 
\end{equation}
(in the last step, we used \Ref{I.2.46.i}). 

\item[(c)]
Eq.\ (\ONE.3.41) (with depth $P \le R$) implies that 
\begin{equation}\label{dble}
\abs{D_h A_v}_0 \leco \abs{h}_0 . 
\end{equation}

\item[(d)]
Eq.\ (\ONE.3.42) again implies \Ref{dble}. 
\end{description}
Thus in all cases, the derivative produces at most an additional factor
$\leco M^{-j_\ell}$ in the bounds.
Applying \Ref{e67}, \Ref{e70} with $w=0$, 
and \Ref{standardCj} with $|\al|=0$, 
counting up factors $M^{j_\ell}$, 
now implies the bound.
\end{proof}

\noin
Thus the effect of a derivative with respect to the dispersion relation
acting on the propagator $S_{\ell,j_\ell}$ 
can be bounded in exactly the same way as 
a derivative with respect to momentum (see Lemma \ref{strider}),
except that $\Impeps $ (which was never actually used) 
has been replaced by zero. 
By Lemma \ref{graphlem} we can again prevent the at most two derivatives 
that appear in the norms from acting on $D_h S_{\ell,j_\ell}$. 
Thus, repeating the argument from \Ref{e123} to \Ref{e124},
again using Lemma \ONE.2.4 and Remark \ONE.2.5, and
using $|j|^R \leco M^{-0.1 j}$, we have
\begin{eqnarray}
\norm{\cY_1(s)}_1 &\leco& M^{-1.1 j} \abs{e_1-e_0}_0,
\\
\norm{\cY_1(s)}_2 &\leco& M^{-2.1 j} \abs{e_1-e_0}_0.
\end{eqnarray}
\intremark{
If the graph has only one external vertex, no shuffling of 
derivatives is needed because at most two derivatives wrt the external
momentum act on the `external' vertex function.
}
Summing the seminorms $\norm{\;\cdot\;}_k$, we get 
\Ref{einbo} and \Ref{zwobo}.

\section{Discussion}\label{dissect}
In this section, we briefly discuss the role of the 
various hypotheses we used in our proofs, 
to summarize which parts of our argument extend
easily to general Fermi surface geometries
and where more work is needed. We also discuss 
the role of the symmetry condition $e(-\Sp) = e(\Sp)$
because cases where this symmetry does not hold are 
interesting from a physical point of view.  

The two main ingredients for the iteration
by which we construct the solution to \Ref{WltfRml}
are
\begin{enumerate}

\item
the existence of an invariant set for the map 
$e \mapsto e+ K^{(R)}$,

\item
the contraction--like bounds \Ref{dr0}, \Ref{dr1}, and \Ref{eq35}.

\end{enumerate}
To prove item 2, we needed only rather weak hypotheses 
on the Fermi surface geometry. In particular, 
we neither used a symmetry $e(-\Sp) = e(\Sp)$
in that part of the proof, nor any assumption about strict 
convexity, nor that $S_e \subset \cF_2$. 
With a different localization operator, defined as in 
\cite{FKST}, one can even drop {\bf F3} 
in the proof of \Ref{einbo} and \Ref{zwobo}
(recall that these bounds imply \Ref{dr1} and \Ref{eq35} by
Theorem \ref{skalensatz} and Lemma \ref{Fgood}). 
However, 
{\bf F3} is also necessary for the Lipschitz continuity,
eq.\ \Ref{dr0}, in 
$\abs{\;\cdot\;}_0$, proven in \ONE, which is essential for 
our iteration estimates. 
One should also keep in mind that if {\bf F3}
does not hold, the selfenergy $\Si$ and the function $K$ will 
in general not even be $C^1$ (in one dimension, where there
are no curvature effects, $\Si$ is not $C^1$; 
this is the source of 
anomalous decay exponents of the two--point function). 

The result that requires the most restrictive hypotheses
is that, for $e \in \Esset$, 
the bound \Ref{r1} for $\abs{K^{(R)}}_2$ holds.
This provides an invariant set for the iteration.
The proof of \Ref{r1}, contained in \TWO\ and \THR, 
uses very detailed geometric estimates which require
convexity and positive curvature, as well as the 
condition $S_e \subset \cF_2$. 

The conditions stated in Section \ref{dispclassect}
(including, in particular, the symmetry (Sy):
$e(-\Sp) = e(\Sp)$ for all $\Sp$)
imply hypotheses (H2)$_{2,0}$, (H3), (H4), and (H5)
of \TWO\ and thus imply \Ref{r1}. In the 
{\em asymmetric case}, where the condition 
(Sy) is dropped, the regularity proof of \TWO\ and \THR\
requires an additional hypothesis, stated as (H4') in \TWO,
which imposes a minimal rate 
of change of the curvature of the Fermi surface 
at those points where the curvature coincides 
with that at the antipode. 
This condition (H4') is not stable under an iteration in 
$\abs{\;\cdot\;}_{3,r}$. 
It is, however, only needed to estimate 
the contributions to $K$ of a very special class of graphs
(the so--called wicked ladders; see Section \TWO.4).
We shall analyze these contributions in a further paper, 
to extend our regularity proof, and thus the inversion theorem, 
to the asymmetric case.
The asymmetry plays a critical role in the proof of the existence 
of a two--dimensional Fermi liquid at zero temperature 
that was announced in \cite{FKLT}. 

The set $\Esset (\dez,\gz,\Gz,\waz) \cap C^3(\cB,\bR)$
of starting $E$ allowed in Theorem \ref{invsatz} is not an 
open subset of $C^2_s (\cB,\bR)$.
However, a look at the more detailed Theorem \ref{deltasatz}
shows that the inversion map really maps the 
ball $\radiEset$, defined in \Ref{radidef}, 
which is open in $\abs{ \cdot }_{3,r}$, to itself
(see \Ref{Ala}). 
Thus in the space of functions with 
bounded radial derivatives, there is an open set 
for which the inversion equation has a solution.
Observe that, for our inversion theorem, 
in contrast to the KAM theorem, there is no 
diophantine condition for irrationality of frequencies. 

As mentioned above, we needed the norm $\abs{\;\cdot\;}_{3,r}$
instead of $\abs{\;\cdot\;}_{2}$ merely for apparently 
rather technical reasons. 
A superficial look at part 4 of Theorem \ref{deltasatz}
even seems to suggest 
that one can extend the inversion map to balls in $\Esset$
that are open in $\abs{\;\cdot\;}_{2}$
However, this is not the case because $\laR$ depends on 
$G_3$, so \Ref{eq39} does not imply that the
inverse map is defined on a dense subset of $B^{(2)}_{\veps/2}$.

\appendix

\section{Proof of Lemma \ref{convsurface}}\label{convexproof}
We first show that \Ref{convangle} follows from \Ref{convradius}.
Fix any $\Sp\in S$. Let $T$ be the tangent plane to $S$ at  $\Sp$ and let
$\Sx$ be the point of $T$ nearest $\Sc$. Since $S$ is convex it lies
on one side of $T$. So the sphere of radius $\sfrac{1}{K}$ centered on
$\Sc$, which by \Ref{convradius} is inside $S$, also lies on one side
of $T$. Hence $\|\Sx-\Sc\|\ge\sfrac{1}{K}$. The vector $\Sx-\Sc$ is 
normal to $T$ and hence parallel to $\Sn(\Sp)$. So $\th(\Sp)$ is 
the angle between $\Sx-\Sc$ and $\Sp-\Sc$ and 
\begin{equation}
\cos\th(\Sp)=\sfrac{\|\Sx-\Sc\|}{\|\Sp-\Sc\|}\ge \sfrac{1/K}{1/k}=\sfrac{k}{K}
\end{equation}

\centerline{\figput{fig1}}

We now prove \Ref{convradius}, starting with $\|\Sp-\Sc\|\ge \sfrac{1}{K}$. 
This is a variant of a classical result.
See, for example, \S24 of \cite{B}. Let $L>K$ and define, for each $\Sp\in S$,
\begin{equation}
\tilde\Sp(\Sp)=\Sp-\sfrac{1}{L}\Sn(\Sp)
\end{equation}
Set
\begin{equation}
\tilde S=\big\{\tilde\Sp(\Sp)\ :\ \Sp\in S\big\}
\end{equation}
Then $\tilde S$ is a $C^1$ surface. 

We claim further that $\Sn(\Sp)$ is normal to $\tilde S$ at 
$\tilde\Sp(\Sp)$. To see this, let $\St$ be a unit
vector that is a principal direction for $S$ at $\Sp$. Call the corresponding
principal curvature $\ka$. Let $\Sq(s)$ be a curve on $S$ that is parametrized
by arc length, passes through $\Sp$ at $s=0$ and has tangent vector 
$\St$ there. Then $s\mapsto \tilde \Sp\big(\Sq(s)\big)=\Sq(s)-
\sfrac{1}{L}\Sn\big(\Sq(s)\big)$ is a curve on $\tilde S$ that passes
through $\tilde\Sp(\Sp)$ at $s=0$ and has tangent vector
\begin{equation}
\sfrac{d\hfill}{ds}\tilde \Sp\big(\Sq(s)\big)\Big|_{s=0}
=\St-\sfrac{1}{L}\sfrac{d\hfill}{ds} \Sn\big(\Sq(s)\big)\Big|_{s=0}
=\St-\sfrac{\ka}{L}\St
\end{equation}
there. Since $\ka<L$, $\St$ is also a tangent vector to $\tilde
S$ at $\tilde\Sp(\Sp)$. As this is the case for all principal directions
$\St$, the tangent plane to $\tilde S$ at $\tilde\Sp(\Sp)$ is parallel
to the tangent plane to $S$ at $\Sp$. 

Since $S$ is strictly convex, with
principal curvatures bounded away from zero, 
 the Gauss map $\Sp\in S\mapsto\Sn(\Sp)$ is bijective and has a $C^1$ inverse
$\Sn\in S^{d-1}\mapsto\Sp(\Sn)\in S$. The map 
$\Sn\in S^{d-1}\mapsto\tilde\Sp\big(\Sp(\Sn)\big)$ is then $C^1$ and surjective.
Furthermore, the normal to $\tilde S$ at $\tilde\Sp\big(\Sp(\Sn)\big)$
is the same as the normal to $S$ at $\Sp(\Sn)$, which is $\Sn$.
Consequently, $\tilde S$ is convex.

As the chord $\Sc_1-\Sc_2$ is of maximal length, it must be parallel to
both $\Sn(\Sc_1)$ and $\Sn(\Sc_2)$. Thus
\begin{equation}
\Sn(\Sc_1)=\frac{\Sc_1-\Sc_2}{\|\Sc_1-\Sc_2\|}=-\Sn(\Sc_2)
\end{equation} 
so that
\begin{equation}
\Sc=\sfrac{1}{2}\big(\Sc_1+\Sc_2\big)
=\sfrac{1}{2}\big(\Sc_1-\sfrac{1}{L}\Sn(\Sc_1)\big)
+\sfrac{1}{2}\big(\Sc_2-\sfrac{1}{L}\Sn(\Sc_2)\big)
\end{equation}
is also the midpoint of a line joining two points of $\tilde S$. By
convexity, $\Sc$ is inside $\tilde S$. The convexity of $\tilde S$ also 
 implies that $\tilde S$ lies
on one side of the tangent plane at $\tilde\Sp(\Sp)$, the side opposite
$\Sn(\Sp)$. Hence $\Sc$, which is inside $\tilde S$ and $\Sp\in S$
are on opposite sides of the tangent plane to $\tilde \Sp(\Sp)$. In particular, the straight line from $\Sc$ to the nearest point, say
$\Sp_0$, of $S$ is parallel to $\Sn(\Sp_0)$ and coincides, in part, with the line from $\tilde \Sp(\Sp_0)$
to $\Sp_0$, which is of length $\sfrac{1}{L}$. We conclude that
$\|\Sp-\Sc\|\ge\sfrac{1}{L}$ for every $L>K$ and every 
$\Sp\in S$. 

\centerline{\figput{fig2}}

The proof that $\|\Sp\|\le\sfrac{1}{k}$ is similar. This time, one lets
$\ell<k$ and defines
\begin{equation}
\tilde\Sp(\Sp)=\Sp-\sfrac{1}{\ell}\Sn(\Sp)
\end{equation}
and sets
\begin{equation}
\tilde S=\big\{\tilde\Sp(\Sp)\ :\ \Sp\in S\big\}
\end{equation}
This time, $\tilde S$, and hence $\Sc$, lies
on the same side of the tangent plane at $\tilde\Sp(\Sp)$ as
$\Sn(\Sp)$. So the straight line from $\Sc$ to the farthest point, say
$\Sp_0$, of $S$ is contained in the line from $\tilde \Sp(\Sp_0)$
to $\Sp_0$, which is of length $\sfrac{1}{\ell}$. 

\centerline{\figput{fig3}}

When $S$ is invariant under inversion in the origin, $\Sn(\Sc_1)=-\Sn(\Sc_2)$
implies that $\Sc_1=-\Sc_2$ so that $\Sc=\half\big(\Sc_1+\Sc_2\big)=\Szero$.

\section{Proof of Lemma \ref{radderiv}}\label{radproof}
Let $\Sp$ be any point of $S_\estart$ and let $\St$ be any
principal direction for $S_\estart$ at $\Sp$. Let $\Sq(s)$ be a
curve on $S_\estart$ that is parametrized by arc length, passes
through $\Sp$ at $s=0$ and has tangent vector $\St$ there. The
principal curvature $\ka$ corresponding to $\St$ obeys
\begin{equation}
\ka\St=\left.\frac{d\hfill}{ds}\frac{\nabla \estart
\big(\Sq(s)\big)} {\|\nabla \estart
\big(\Sq(s)\big)\|}\right|_{s=0}
=\frac{\estart''(\Sp)\St}{\|\nabla \estart(\Sp)\|}+ \nabla \estart
\big(\Sp\big)\left.\frac{d\hfill}{ds}\frac{1} {\|\nabla \estart
\big(\Sq(s)\big)\|}\right|_{s=0}
\end{equation}
and hence
\begin{equation}
\ka=\frac{\big(\St,\estart''(\Sp)\St\big)}{\|\nabla
\estart(\Sp)\|}
\end{equation}
Consequently, $S_\estart$ is a convex surface that is invariant
under inversion in the origin and has all principal curvatures
between $\sfrac{\waz}{\Gz}$ and $\sfrac{\Gz}{\gz}$. By Lemma
\ref{convsurface},
\begin{equation}
\frac{\del}{\del r} \estart(\pp(r,\th)) = \nabla \estart
(\pp(r,\th)) \cdot \frac{\del\Sp}{\del r} (r,\th) \ge \|\nabla
\estart (\pp(r,\th))\|\frac{\waz/\Gz}{\Gz/\gz} \ge
\frac{\waz\gz^2}{\Gz^2}
\end{equation}
for all $r=\rF(\estart,\th)$. Choose
$g_1=\frac{\waz\gz^2}{4\Gz^2}$ and 
$r_0=\min\{\frac{g_1}{\Gz}, \dez\}$. Then
\begin{eqnarray}
\frac{\del}{\del r} \estart(\pp(r,\th)) &=&\nabla \estart
\big(\pp(\rF(\estart,\th),\th)\big) \cdot \sfrac{\del\pp}{\del r}
(r,\th)\\ & &+\Big[\nabla \estart \big(\pp(r,\th)\big)
-\nabla \estart \big(\pp(\rF(\estart,\th),\th)\big) \Big]\cdot
\sfrac{\del\pp}{\del r} (r,\th)\nonu
\end{eqnarray}
and
\begin{equation}
\Big|\Big[\nabla \estart \big(\pp(r,\th)\big) -\nabla \estart
\big(\pp(\rF(\estart,\th),\th)\big) \Big]\cdot
\sfrac{\del\pp}{\del r} (r,\th)\Big| \le
\Gz\big|r-\rF(\estart,\th)\big|
\end{equation}
so
\begin{equation}
\sfrac{\del}{\del r} \estart(\pp(r,\th))\ge 2g_1 \qquad\mbox{ for
all }\ \ \big|r-\rF(\estart,\th)\big|\le 2r_0,\ \th\in S^{d-1}
\end{equation}
Similarly, if $\abs{e-\estart}_1\le g_1$,
\begin{equation}
\sfrac{\del}{\del r} e(\pp(r,\th))\ge g_1 \qquad\mbox{ for all }\
\ \big|r-\rF(\estart,\th)\big|\le 2r_0,\ \th\in S^{d-1}
\end{equation}
This verifies \Ref{rg0}. We merely need to choose $\veps<g_1$.

To verify \Ref{pfs}, observe that if $\abs{e-\estart}_0\le
r_0g_1$, then $\abs{e\big(\rF(\estart,\th),\th\big)}\le r_0g_1$
and hence
\begin{equation}
\big|\rF(e,\th)-\rF(\estart,\th)\big|\le r_0
\end{equation}
by \Ref{rg0}.

The same argument that shows that $\Esset$ is open in
$(C_s^2(\cB,\bR), \abs{\;\cdot\;}_2)$ also yields $e\in\Esset
(\dez/2,\gz/2,2\Gz,\waz/2 )$, if we choose $\veps$ small enough,
depending only on $\dez,\ \gz,\ \Gz$ and $\waz$.

\newpage

\section*{Acknowledgements}

We thank H. Kn\"orrer for suggesting the proof of Lemma
\ref{convsurface}.


\begin{thebibliography}{1}
\bibitem{B} W.\ Blaschke, {\sl Kreis und Kugel}, Verlag von Veit \& Comp.,
Leipzig, 1916.
\bibitem{FST1} J.\ Feldman, M.\ Salmhofer, and E.\ Trubowitz,
{\sl J.\ Stat.\ Phys. \bf 84} (1996) 1209--1336
\bibitem{FST2} J.\ Feldman, M.\ Salmhofer, and E.\ Trubowitz,
{\sl Comm.\ Pure Appl.\ Math. \bf LI} (1998) 1133--1246
\bibitem{FST3} J.\ Feldman, M.\ Salmhofer, and E.\ Trubowitz,
{\sl Comm.\ Pure Appl.\ Math. \bf LII} (1999) 273--324
\bibitem{S}M.\ Salmhofer,
{\sl Rev.\ Math.\ Phys.\ \bf 10} (1998) 553--578
\bibitem{crg} M.\ Salmhofer, {\sl Commun.\ Math.\ Phys.\
\bf 194} (1998) 249--295
\bibitem{FMRT} J.\ Feldman, J.\ Magnen, V.\ Rivasseau, and E.\ Trubowitz,
{\sl Helv.\ Phys.\ Acta \bf 65} (1992) 679--721
\bibitem{Brisbane}  J.\ Feldman, M.\ Salmhofer, and E.\ Trubowitz,
Renormalization of the Fermi Surface, to appear in {\it $XII^{\rm
th}$ International Congress of Mathematical Physics}.
\bibitem{FKLT}J.\ Feldman, H.\ Kn\" orrer, D.\ Lehmann, E.\ Trubowitz,
in {\sl Constructive Physics},
V. Rivasseau (ed.), Springer Lecture Notes in Physics, 1995
\bibitem{FKST}J.\ Feldman, H.\ Kn\" orrer, M.\ Salmhofer, E.\ Trubowitz,
{\sl J.\ Stat.\ Phys.\ \bf 94} (1999) 113--157

\end{thebibliography}
\end{document}